\documentclass[11pt,a4paper]{article}

\usepackage[utf8]{inputenc}
\usepackage{amsmath, amsfonts, amssymb, mathtools}

\usepackage[a4paper, left=1in, right=1in]{geometry}
\usepackage{fancyhdr}
\allowdisplaybreaks[1]

\usepackage{booktabs}
\usepackage{array}
\usepackage{float}
\usepackage{graphicx}
\usepackage[margin=10pt, font=small, labelfont=bf, labelsep=endash]{caption}

\usepackage{appendix}

\usepackage[
backend=biber,
style=apa,
bibstyle=authoryear,
citestyle=authoryear,
maxcitenames=2,
maxbibnames=99
]{biblatex}
\DeclareSourcemap{
  \maps[datatype=bibtex]{
    \map{
      \step[fieldsource=doi, final]
      \step[fieldset=url, null]
    }
  }
}

\usepackage{hyperref}
\hypersetup{
    pdftitle={Seasonal Trading in Commodity Futures: Evidence from Regression and Singular Spectrum Signals},
    pdfauthor={Ralph Kosch and Robin Forsberg},
    pdfsubject={Commodity futures seasonality, out-of-sample trading strategies, and statistical robustness},
    pdfkeywords={commodity futures, seasonality, dummy-variable regression, Singular Spectrum Analysis, robust Singular Spectrum Analysis, out-of-sample trading, Maximum Entropy Bootstrap}
}
\usepackage[nameinlink,capitalize]{cleveref}

\newcommand{\keywords}[1]{\par\vspace{0.75em}\noindent\textbf{Keywords:} #1}

\title{Seasonal Trading in Commodity Futures:\\
Evidence from Regression and Singular Spectrum Signals}

\author{%
\begin{tabular}{c}
Ralph Kosch\textsuperscript{1}
\quad \& \quad
Robin Forsberg\textsuperscript{1,2}\\[0.6em]
{\footnotesize
\texttt{ralphpeter.kosch@uzh.ch}
\qquad
\texttt{robin.forsberg@helsinki.fi}}\\[0.45em]
{\small \textsuperscript{1}University of Zurich, Switzerland}\\[0.1em]
{\small \textsuperscript{2}University of Helsinki, Finland}
\end{tabular}%
}

\date{September 2026}

\begin{document}
\maketitle

\begin{abstract}
Commodity futures are shaped by harvest cycles, weather shocks, storage conditions, and
seasonal demand, but it remains unclear whether recurring patterns yield robust out-of-sample
trading profits. Existing research documents return seasonality in commodity futures as well as
more complex seasonal structure, while leaving less evidence on how alternative seasonal models
compare under common implementation constraints. This article compares dummy-variable
regression (DVR), Singular Spectrum Analysis (SSA), and robust low-rank SSA (RLSSA) within a
unified trading framework, including a volatility-normalised specification. Using monthly
delivery-avoidance returns for 15 liquid commodity futures, the models are estimated on rolling
ten-year windows and evaluated from 2016 to 2024 with transaction costs, an equal-weight long
benchmark, and Maximum Entropy Bootstrap (MEB) assessment. Across 500 MEB paths, the
benchmark has the strongest average full-period profile, with a cumulative return of 16.81\%,
a Sharpe ratio of 0.191, and a maximum drawdown of -0.414. Classical SSA has the strongest
average model outcomes, but negative median cumulative returns and deep drawdowns indicate
substantial path sensitivity. Volatility normalisation reduces the average DVR short loss but
generally weakens SSA-based portfolios. None of the 18 approximate paired MEB Sharpe tests
rejects after within-family Holm adjustment. Because MEB preserves each contract's temporal
rank ordering, the assessment is conditional on observed timing rather than a timing-randomised
seasonal null. The evidence does not establish robust benchmark outperformance and shows that
model performance varies materially across market subperiods.
\end{abstract}

\keywords{commodity futures, seasonality, dummy-variable regression, singular spectrum analysis, robust singular spectrum analysis, out-of-sample trading, bootstrap}

\newpage

\section{Introduction}

Seasonality in commodity futures is rooted in the physical economy. Harvest cycles, planting decisions, storage dynamics, weather shocks, extraction schedules, transport bottlenecks, and seasonal demand patterns all shape the timing of supply and demand. As a result, commodity returns may display recurring calendar structure. At the same time, these patterns need not be fixed. Seasonal effects can shift across regimes, weaken or strengthen after major shocks, and emerge at frequencies shorter than the standard annual cycle. This makes commodity seasonality an attractive but difficult object of research, since the challenge is not only to detect recurring patterns, but to determine whether they survive in implementable out-of-sample trading strategies.

The existing literature \parencite{DEGENHARDT2018169, TangEtAl2024, Rodrigues2018RobustSSA} suggests that three methodological approaches are especially relevant. First, dummy-variable regression provides the most transparent way to test calendar effects by estimating whether returns in a given month differ from returns in the rest of the year. \Textcite{DEGENHARDT2018169} revisit the Sell in May effect across equities and commodities. They analyse highly liquid commodity futures represented by the aggregate GSCI and its futures-based sub-indices and find that the evidence and practical value of the effect vary across markets and specifications. This is important because it shows that simple calendar-based regressions can detect heterogeneous seasonal anomalies in commodity futures without implying a uniform aggregate premium.

Second, Singular Spectrum Analysis (SSA) offers a more flexible and data-driven alternative when seasonalities are not limited to one fixed annual pattern. Rather than imposing a pre-specified structure, SSA decomposes a time series into trend, oscillatory components, and noise. In a commodity setting, this flexibility matters because harvest cycles, weather variation, inventory dynamics, and market stress can generate overlapping quarterly, semi-annual, and annual effects. \Textcite{TangEtAl2024} analyse 35 commodity futures from 2000 to 2022 across multiple temporal aggregation levels and show that SSA identifies complex and heterogeneous seasonal structure whose patterns vary with market uncertainty. Their evidence suggests that commodity seasonality can be multi-frequency and time-varying, which makes SSA a natural candidate for a broader investigation of seasonal trading signals.

Third, robust $L_1$ singular spectrum analysis (RLSSA) extends classical SSA to settings in which outliers can distort the least-squares singular value decomposition. Commodity futures are repeatedly exposed to extreme observations created by supply disruptions, geopolitical shocks, storage squeezes, and abrupt reversals in investor positioning. \Textcite{Rodrigues2018RobustSSA} develop a robust SSA approach for contaminated series, while \textcite{HawkinsLiuYoung2001} provide the alternating $L_1$ singular value decomposition used in the present implementation. This motivates examining whether robust low-rank extraction changes the trading signals in commodity returns with heavy tails and break-like episodes. It does not presume that robustification must improve portfolio performance.

The literature establishes that seasonality in commodity futures is economically plausible and methodologically measurable, while also showing that return seasonality is not uniformly persistent through time. \Textcite{KeloharjuLinnainmaaNyberg2016} document same-calendar-month return seasonalities across several asset classes, including commodities. More directly, \textcite{LiLiuMiaoTse2024} examine monthly returns for 26 commodity futures from 1970 to 2023 and find that seasonal effects have weakened in more recent periods. \Textcite{DEGENHARDT2018169} likewise show that the evidence and practical value of a specific calendar rule vary across markets and specifications. These studies establish a direct literature on return seasonality in commodity markets. What remains less explored is whether transparent calendar regression and more flexible SSA-based approaches yield economically useful signals when evaluated side by side under the same implementation framework. The SSA and RLSSA literatures focus primarily on decomposition quality and forecasting performance rather than on a common out-of-sample trading design incorporating delivery avoidance, liquidity screening, transaction costs, and benchmark comparisons \parencite{TangEtAl2024, Rodrigues2018RobustSSA, KazemiRodrigues2025}. The present study addresses this intersection by evaluating regression-based seasonality, classical SSA, and robust SSA under identical rolling estimation and portfolio rules. This distinction matters because detecting seasonality in returns is not the same as trading it. A pattern can be statistically visible in historical data but still fail once it must be implemented through liquid contracts, repositioned through time, traded after costs, and compared with a simple benchmark.

This article addresses that gap. It studies whether systematic seasonality models can turn recurring structure in commodity futures into robust excess returns relative to a transparent equal-weight long benchmark, and whether a specification based on volatility-normalised returns improves selection stability and portfolio performance. More specifically, the study tests two linked propositions. First, long and short seasonality strategies based on DVR, SSA, and RLSSA may generate out-of-sample performance that exceeds a simple passive long-term commodity strategy. Second, volatility normalisation may improve the comparability of contracts and lead to more stable strategy outcomes. The same long benchmark is retained deliberately on the short side because the research question is whether active short selection can beat that conventional long-only reference, rather than whether it can beat passive short exposure. Beyond these tests, the article contributes a single transparent empirical framework that can serve as a baseline for future research on commodity seasonality.

Using monthly delivery-avoidance returns for 15 liquid commodity futures markets from 2006 to 2024, with seasonal signals estimated on rolling ten-year windows and evaluated out of sample from 2016 to 2024, the analysis shows that exploitable seasonality depends strongly on the combination of model and direction and is much less stable than the raw seasonal intuition would suggest. None of the long- or short-side model comparisons rejects against the equal-weight long benchmark after the within-family Holm adjustment. Classical SSA provides the strongest model-based average outcomes, but positive means coexist with negative medians and deep drawdowns. DVR short portfolios lose money on average under both specifications, and RLSSA does not consistently improve on classical SSA under the fixed \(r=12\) design. The strongest gains are concentrated in particular subperiods rather than sustained across the evaluation window. These findings matter for both research and practice. For researchers, they show that the conversion from signal detection to benchmark outperformance is fragile. For institutional investors, the results point less to an immediately implementable strategy than to the need for stricter signal filters, contract-specific calibration, and more conservative robustness testing.

To reach these results in a structured way, the next section presents the data construction and the seasonality models. Section 3 then explains portfolio formation, benchmark design, performance metrics, and the bootstrap-based statistical evaluation. Section 4 reports the core findings and integrates the main robustness evidence into the results narrative. Section 5 discusses the economic interpretation and implementation realism of the strategies. Section 6 concludes and outlines the broader relevance of the findings for future commodity research.

\section{Data and Methods}

\subsection{Data}

\subsubsection{Commodities}

The empirical analysis is based on monthly futures returns for 15 commodity markets. The final universe comprises cocoa (CC), coffee (CF), WTI crude oil (CO), copper (CP), cotton \#2 (CT), gold (GD), lean hogs (HE), heating oil (HO), live cattle (LE), natural gas (NG), sugar \#11 (SU), silver (SV), corn (ZC), soybeans (ZS), and wheat (ZW). Raw contract histories were obtained online \parencite{tradingcharts}. The commodity universe is screened for liquidity before the out-of-sample evaluation begins, as described in Section~\ref{sec:tradability-filter}. For each retained commodity and calendar month, an eligible contract must expire at least two calendar months after the observation month and contain price coverage through at least 14 days after month-end. The nearest eligible expiry is selected, and the monthly simple return is measured from the first trading day's open to the last trading day's close. Log returns used for estimation are obtained from these simple returns. This construction keeps the sample close to economically tradeable contracts while avoiding nearby expiries that are unsuitable for the intended monthly holding rule. It is a delivery-avoidance construction rather than a literal front-month series.\footnote{This subsection is adapted directly from \textcite{Kosch2025Thesis}. Code and derived monthly panels are available in the \href{https://github.com/RPKosch/Seasonal-Trading-in-Commodity-Markets-Preprint-Version}{public GitHub repository}; the complete 500-run artifacts are available from the authors. The underlying market histories are attributed to their provider.}

\subsubsection{Tradability filter and sample construction}
\label{sec:tradability-filter}

To ensure implementability, the commodity universe is screened using liquidity observed before the out-of-sample evaluation begins. For each candidate commodity, monthly average daily trading volume is evaluated over the January 2006 to December 2015 pre-evaluation period. A commodity is retained only if its minimum monthly average daily volume over this period is at least 1,000 contracts. The resulting commodity universe is therefore fixed before the first out-of-sample month in January 2016 and is not redefined using future out-of-sample liquidity information. The screen uses historically demonstrated liquidity to identify a set of consistently active commodity futures before the trading evaluation begins. In the underlying thesis, preliminary series were first constructed with a lower cutoff of 500 contracts per day, after which the threshold was increased to 1,000 contracts without eliminating any additional usable tickers. The 1,000-contract threshold therefore provides a stricter pre-evaluation tradability screen while preserving a broad cross section of commodity markets.

In an earlier stage of the sample construction, palladium and platinum futures were also considered. Both contracts, however, fell clearly below the required liquidity standard in the early part of the sample and were therefore excluded from the final candidate sample. A complete overview of tickers, contract names, and trading volumes is reported in Appendix Table~\ref{tab:sample-volume}. The table reports liquidity descriptively over the full available panel and does not redefine the commodity universe after the out-of-sample evaluation has begun.

\subsubsection{Contracts and sample period}

The retained monthly panel begins in January 2001. The broader 2001--2024 window is used only for the representative return plots in the next section. Model estimation starts in January 2006, and the final evaluation month is December 2024.

December 2024 is a fixed study endpoint rather than the last observation currently stored in the repository. Any later observations are excluded from the representative plots, rolling estimation windows, realised strategy returns, and bootstrap inputs. Keeping a common endpoint preserves the original research design and prevents post-sample developments from influencing the reported evidence.

The resulting dataset covers several economically distinct sectors, including energy, metals, soft commodities, grains, and livestock. This broad coverage is important because commodity seasonality is unlikely to reflect a single common mechanism. Instead, it may arise from sector-specific factors such as weather exposure, storage conditions, biological production cycles, supply disruptions, and inventory dynamics. A broad panel is therefore needed to assess whether seasonal return patterns are systematic across markets or limited to particular contracts.

With the sample period and sectoral coverage defined, the next section presents a small set of representative return paths to provide an initial descriptive view of the heterogeneity in the data before turning to the formal econometric analysis.

\subsubsection{Representative return profiles}

\begin{figure}[H]
\centering
\includegraphics[width=0.32\linewidth]{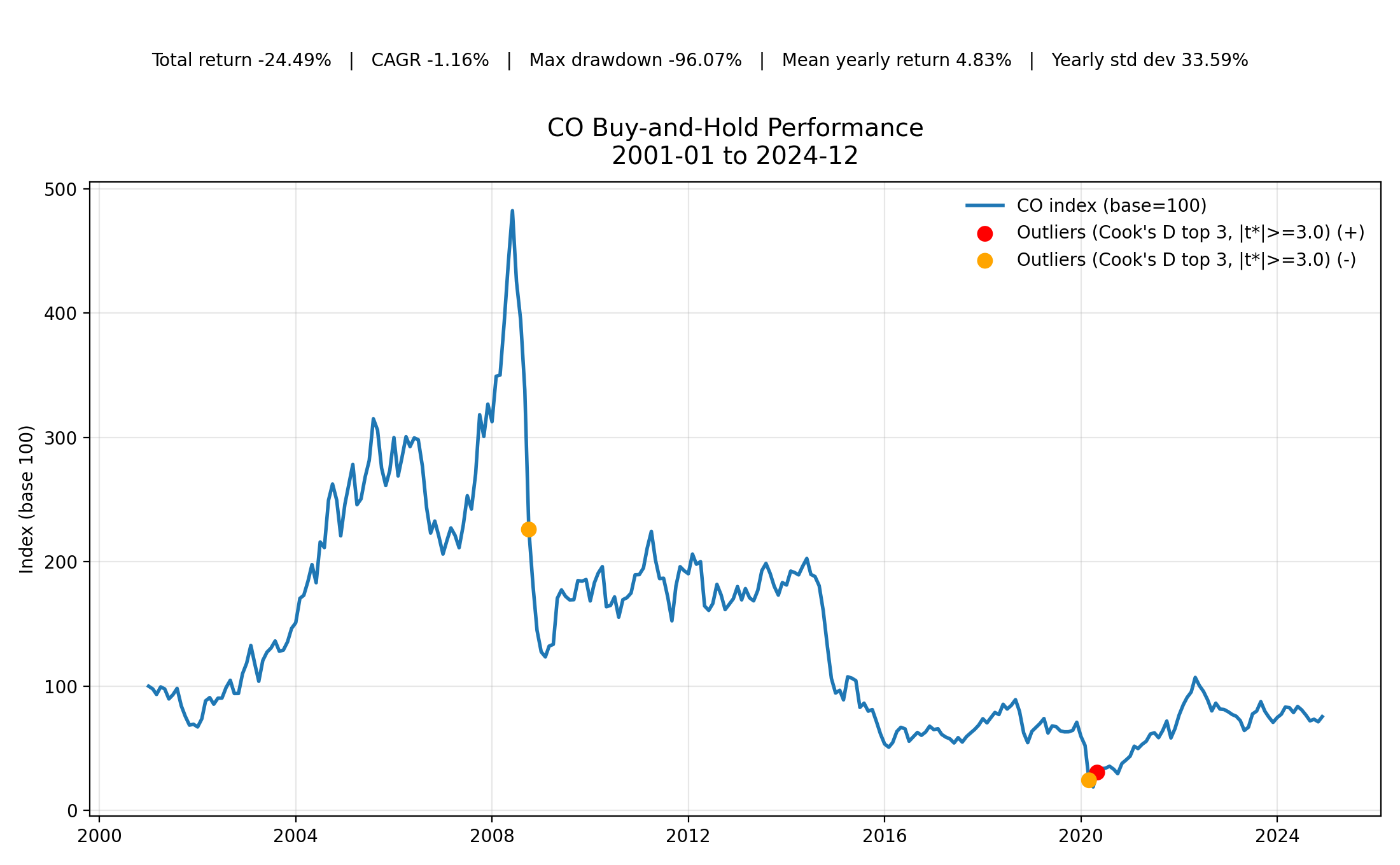}\hfill
\includegraphics[width=0.32\linewidth]{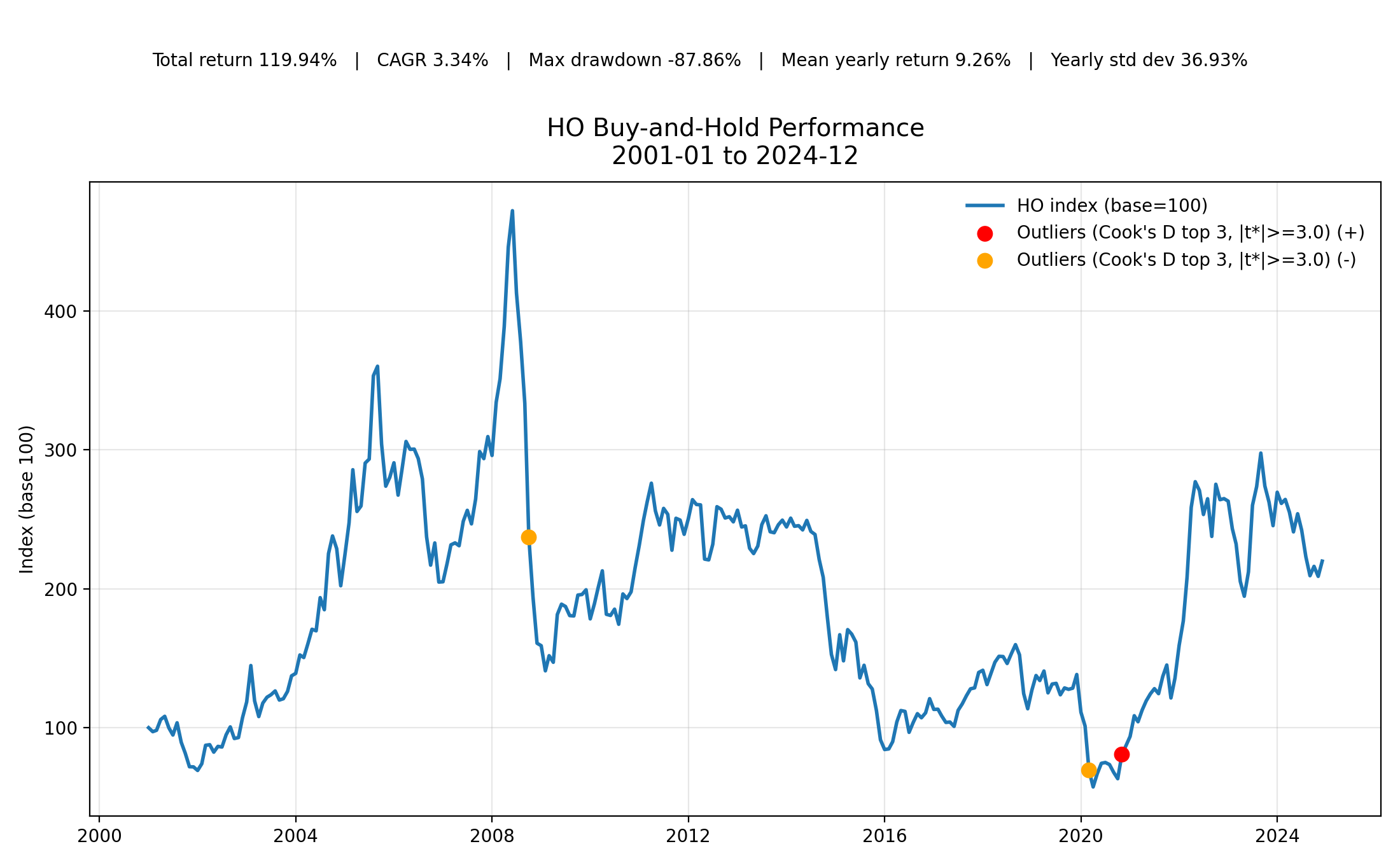}\hfill
\includegraphics[width=0.32\linewidth]{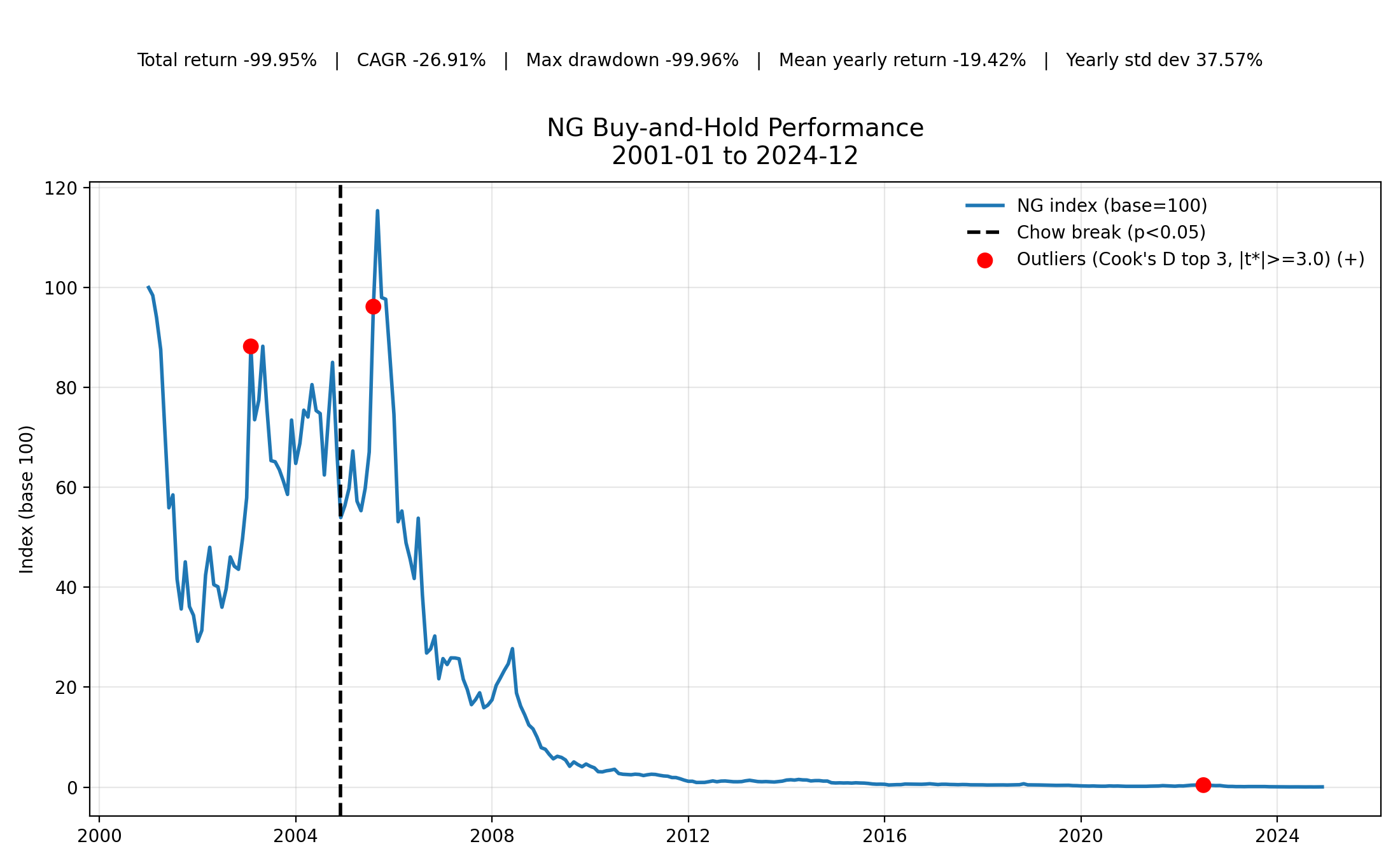}\par\vspace{0.5em}
\includegraphics[width=0.32\linewidth]{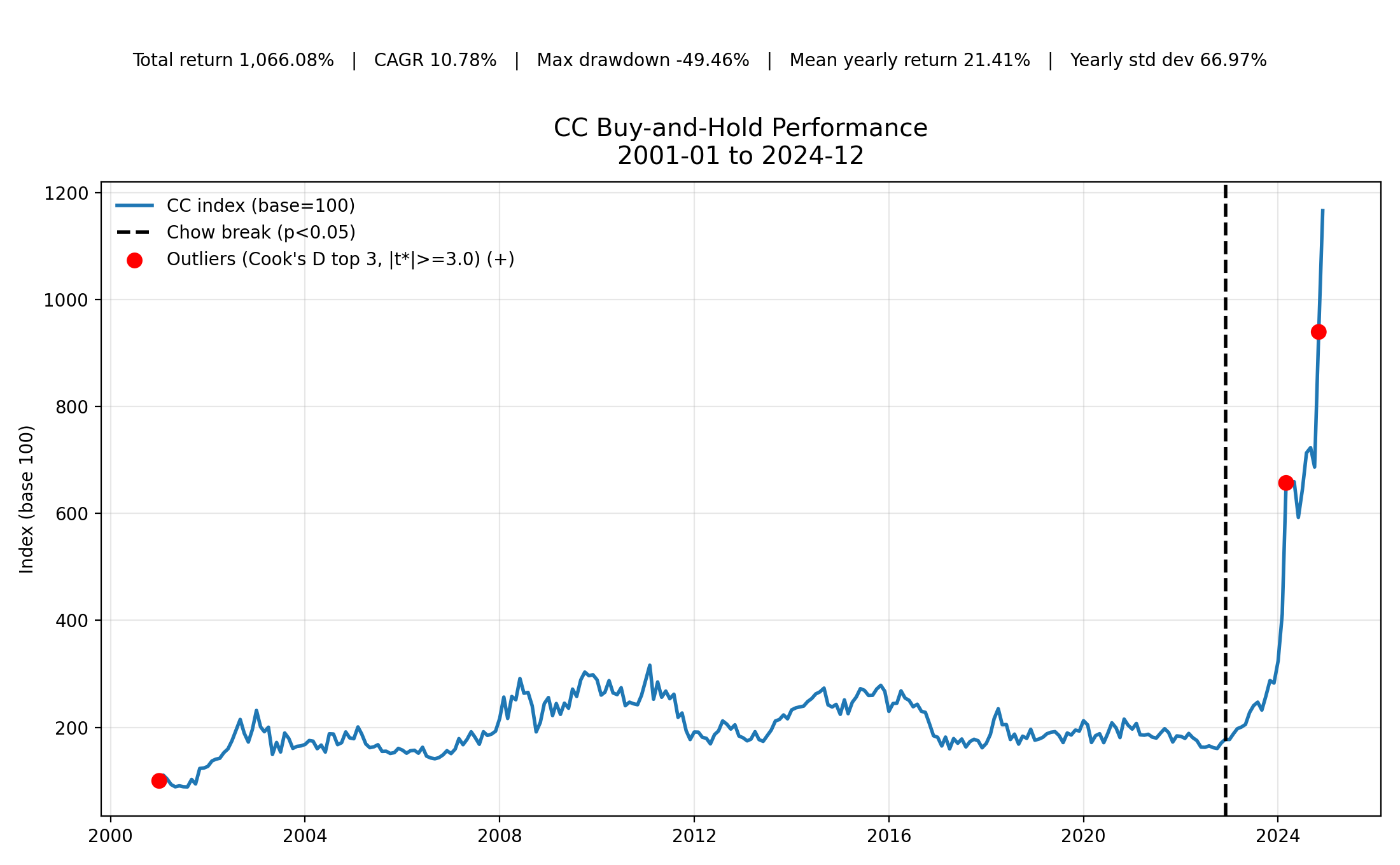}\hfill
\includegraphics[width=0.32\linewidth]{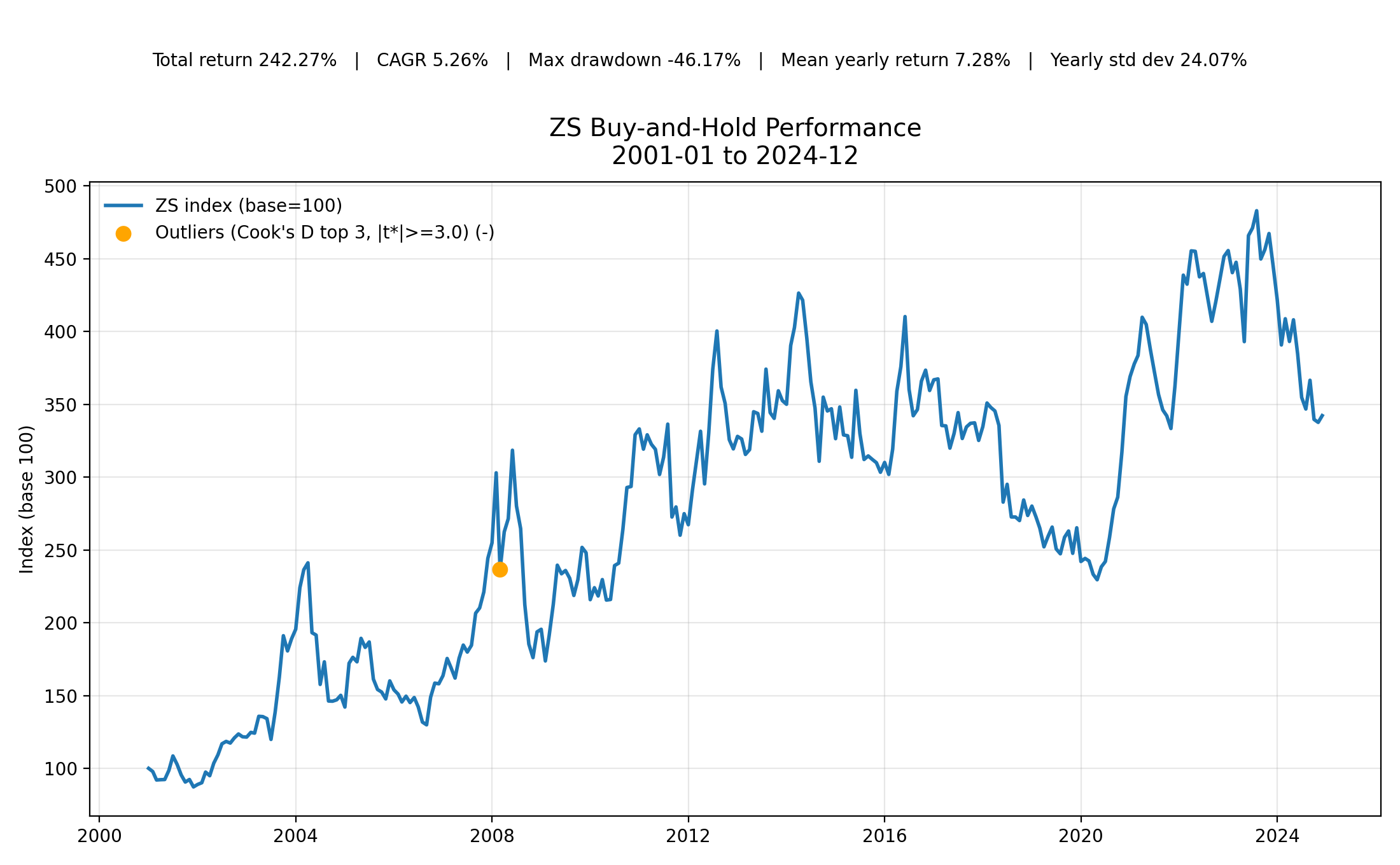}\hfill
\includegraphics[width=0.32\linewidth]{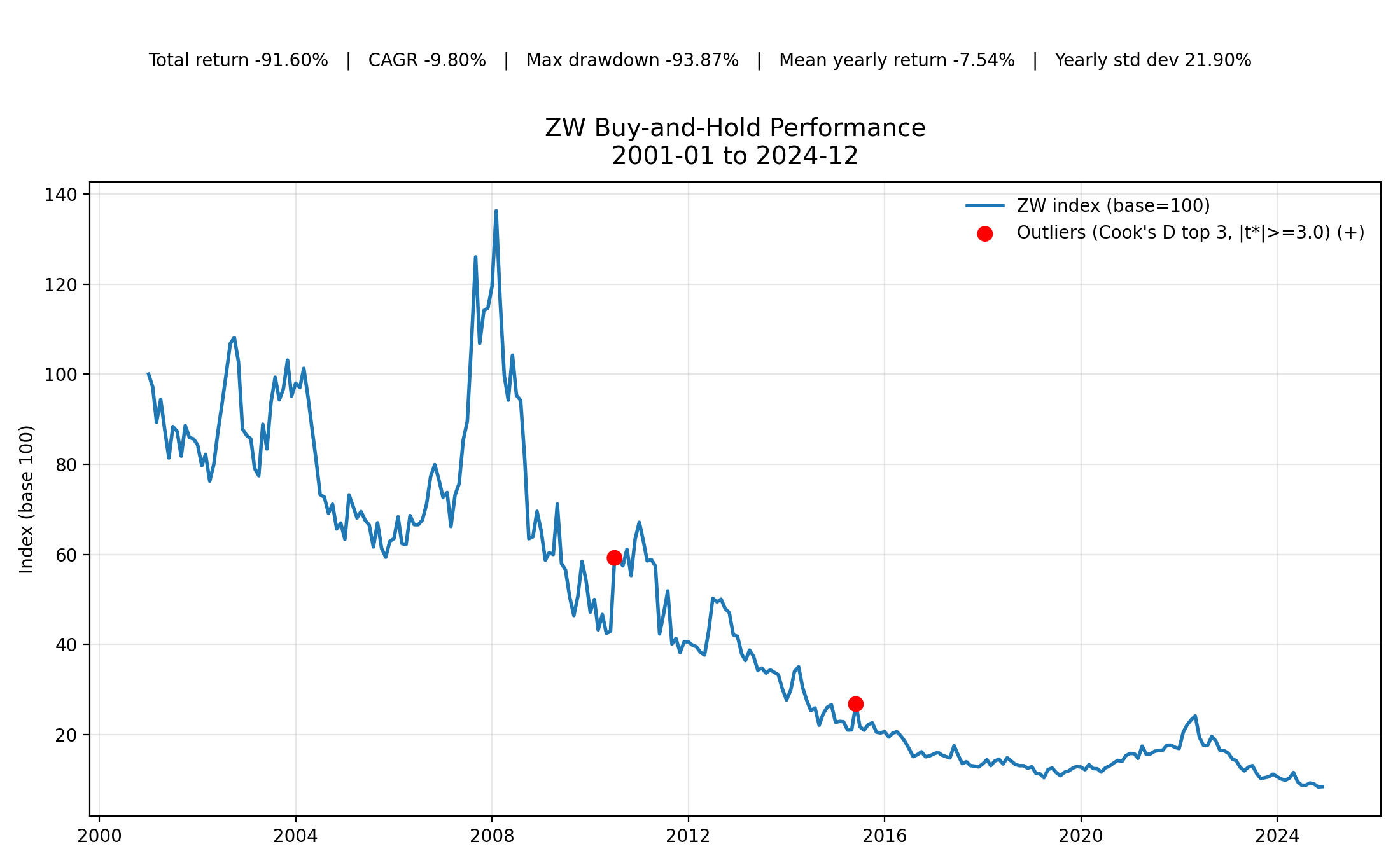}
\caption{Representative cumulative delivery-avoidance return profiles, January 2001--December 2024}
\label{fig:representative_returns}
\end{figure}

Figure \ref{fig:representative_returns} reports six representative cumulative return profiles for WTI crude oil, heating oil, natural gas, cocoa, soybeans, and wheat. Each profile compounds the monthly simple returns from the nearest eligible delivery-avoiding contract and is therefore not a conventional buy-and-hold return on one fixed contract. Its purpose is descriptive: the figure shows that the panel combines markets with very different long-horizon paths, volatilities, drawdowns, and break-like episodes.

Two features of Figure \ref{fig:representative_returns} are particularly informative for understanding the data. First, even closely related contracts can display materially different long-run behaviour. WTI crude oil and heating oil are exposed to similar macroeconomic and energy-market shocks, with both series showing pronounced movements around the 2008 commodity boom and the 2020 collapse. Yet their long-horizon outcomes differ substantially. Heating oil ends the period with a positive total return of 105.62\% and a CAGR of 3.05\%, whereas WTI crude oil ends with a total return of -18.30\% and a CAGR of -0.84\%. This contrast shows that common sector exposure does not imply identical return dynamics at the contract level.

Second, the cross section contains extreme heterogeneity. Natural gas has a particularly poor long-horizon delivery-avoidance profile, ending with a total return of -99.96\%, a CAGR of -28.08\%, and a near-complete drawdown. Cocoa shows the opposite pattern. Its cumulative profile remains comparatively moderate for most of the sample and then accelerates sharply at the end, producing a total return of 1,448.69\% and a CAGR of 12.09\%. This late-sample surge is visually striking and indicates that long-run commodity returns may be shaped not only by gradual cycles but also by concentrated regime shifts and supply-driven episodes. A related contrast appears within grains, where soybeans finish strongly positive with a total return of 208.55\%, while wheat declines persistently and ends with a total return of -91.82\%. Taken together, these series illustrate why the data should be treated as a heterogeneous panel rather than as a uniform commodity block.

\subsubsection{Implications for the empirical analysis}

The descriptive evidence already points to an important feature of the dataset. Commodity futures returns are not characterised by a single common profile. Instead, the panel combines sustained appreciation, prolonged decline, sharp reversals, and highly uneven volatility. Even when recurring patterns are present, they are embedded in a market environment vulnerable to political developments, geopolitical tensions, weather events, and other shocks. The profiles alone cannot establish how these events affect a trading rule, but they motivate an out-of-sample design that measures drawdowns and tests whether apparent seasonal performance is stable across alternative paths and subperiods.

The resulting data construction yields a liquid but highly heterogeneous panel of commodity futures returns. This heterogeneity is central to the empirical design, because the goal is not only to detect recurring patterns within individual contracts, but also to compare seasonality signals across very different markets under a common framework. The next subsection therefore introduces the estimation methods used to generate these signals.

\subsection{Methods}

This section describes how one-step-ahead seasonality signals are estimated for each commodity contract. The emphasis is on signal construction within a common rolling framework and on ensuring that scores are comparable across contracts. Portfolio formation, benchmark construction, and statistical evaluation are discussed in the next section.

\paragraph{Empirical design.}
The analysis follows a strict rolling out-of-sample signal design based on monthly log returns. For each commodity \(i\) and evaluation month \(t\), the model is estimated only on the preceding 120 months of data. This ten-year window is long enough to capture repeated calendar structure while still allowing the estimated signal to adjust over time. The first out-of-sample signal is generated for January 2016 using observations from January 2006 to December 2015, and the procedure is then rolled forward one month at a time until December 2024. All three model classes use the same rolling windows and evaluation horizon. In both the historical application and every synthetic path, the signal for a trading month is estimated without using that month's realised return.

\paragraph{Dummy-variable regression.}
Dummy-variable regression serves as the parametric benchmark for calendar seasonality. For each contract \(i\) and each evaluation month \(t+1\), let \(m\) denote the corresponding calendar month. Within the current rolling estimation window, we estimate
\begin{equation}
r_{i,t} = \alpha_{i,m} + \beta_{i,m} D_{m,t} + \varepsilon_{i,t},
\label{eq:dvr}
\end{equation}
where \(r_{i,t}\) is the monthly log return and \(D_{m,t}\) equals one if observation \(t\) falls in month \(m\) and zero otherwise. In this specification, \(\alpha_{i,m}\) is the average return across all months other than \(m\), while \(\beta_{i,m}\) measures the return differential between month \(m\) and the average of the remaining eleven months within the current estimation window. A positive \(\beta_{i,m}\) therefore indicates that month \(m\) outperforms the rest of the year on average. Inference uses two-sided Newey--West heteroskedasticity- and autocorrelation-consistent standard errors \parencite{NeweyWest1987}, with the Bartlett kernel and the fixed lag rule \(\max\{1,\lfloor0.75T^{1/3}\rfloor\}\), which gives three lags when \(T=120\). The 10 percent threshold is an explicit trading filter: a long candidate must satisfy \(\hat\beta_{i,m}>0\) and \(p_{i,m}\leq0.10\), while a short candidate must satisfy \(\hat\beta_{i,m}<0\) and \(p_{i,m}\leq0.10\). It is used solely as a trading screen and is not interpreted as a cross-sectional discovery test. For trading month \(t+1\), eligible contracts are ranked by \(\hat{\beta}_{i,m}\); if no contract is eligible on a given side, the DVR portfolio remains in cash for that month and earns zero before and after costs. Related calendar-dummy and seasonal-regression treatments are discussed elsewhere \parencite{DEGENHARDT2018169, rozeff1976capital, Cipra2020Seasonality}.

\paragraph{Singular Spectrum Analysis.}
To allow for non-parametric seasonal structure, the study also applies Singular Spectrum Analysis. For each rolling window, let \((x_1,\dots,x_N)\) denote the \(N=120\) monthly log returns of a given contract. With embedding dimension \(L=36\) and \(K=N-L+1\), the series is mapped into the Hankel trajectory matrix
\[
\mathbf{X} =
\begin{pmatrix}
x_1 & x_2 & \cdots & x_K \\
x_2 & x_3 & \cdots & x_{K+1} \\
\vdots & \vdots & \ddots & \vdots \\
x_L & x_{L+1} & \cdots & x_N
\end{pmatrix}.
\]
The trajectory matrix is decomposed by singular value decomposition, the first \(r=12\) components are retained as signal, and the corresponding series is reconstructed by diagonal averaging. A one-step-ahead forecast is then generated by recurrent SSA, yielding the signal \(\widehat r^{\mathrm{SSA}}_{i,t+1}\). The choice \(L=36\) spans three annual cycles and lies close to the \(N/3\) forecasting guideline discussed by \textcite{GolyandinaKorobeynikov2014}. The retained rank \(r=12\) is a fixed design choice applied uniformly to every contract and rolling window for comparability. \Textcite{Hassani2007SSA} provides a monthly SSA application using a 12-eigentriple reconstruction, but that application is not treated as a universal rank-selection rule.

\paragraph{Robust low-rank SSA.}
Because commodity returns often contain jumps, heavy tails, and abrupt regime shifts, the analysis also considers a robust low-rank SSA variant. RLSSA replaces the least-squares decomposition with the Hawkins alternating \(L_1\) (AL1) robust SVD \parencite{HawkinsLiuYoung2001}. The first 12 sequentially deflated components form a robust rank-12 trajectory approximation; diagonal averaging reconstructs the signal, which is then re-embedded as a Hankel matrix. An ordinary SVD of this re-Hankelised signal supplies the orthonormal vectors for the same recurrent one-step forecast used by SSA. The AL1 iteration uses the squared-vector convergence threshold \(10^{-10}\), permits at most 50,000 iterations per component, and raises an error rather than substituting a classical SVD if convergence fails. To isolate robustness from retuning, RLSSA uses the same fixed structural parameters as SSA in every contract and window: \(N=120\), \(L=36\), and \(r=12\). \Textcite{Rodrigues2018RobustSSA} and \textcite{KazemiRodrigues2025} provide the robust-SSA context for this design. More recent multivariate diagonalwise low-rank formulations are related but algorithmically distinct from the univariate AL1 implementation used here \parencite{CentofantiEtAl2025}.

\paragraph{Scaling of raw seasonality forecasts.}
Raw seasonality outputs are not directly comparable across contracts because all three methods inherit the scale of the underlying return series. If two contracts share the same seasonal structure but one exhibits larger return magnitudes, its raw seasonal forecast will also be larger even though the underlying seasonality is not economically stronger. To illustrate this, cocoa futures are used as an example. Monthly log returns from January 2005 to December 2014 are denoted by \(\{r_t\}\), and a scaled series is created by multiplying each observation by a constant \(c\). We compare the original series \((c=1)\) with a rescaled version \((c=2)\). For DVR, the reported quantity is the estimated seasonal coefficient \(\hat\beta_m\). For SSA and RLSSA, the reported quantities are the corresponding one-step-ahead forecasts.

\begin{table}[htbp]
  \centering
  \caption{Raw seasonality outputs scale one-for-one with return magnitude in all three models}
  \label{tab:scaling_experiment}
  \begin{tabular}{lccc}
    \toprule
    & DVR \(\hat\beta_m\) & SSA forecast & RLSSA forecast \\
    \midrule
    \(c = 1\) & \(0.0299894\)  & \(-0.0192349\) & \(0.0148370\) \\
    \(c = 2\) & \(0.0599787\)  & \(-0.0384698\) & \(0.0296740\) \\
    \bottomrule
  \end{tabular}
\end{table}

Table~\ref{tab:scaling_experiment} shows that doubling the return series approximately doubles the DVR coefficient and the SSA and RLSSA forecasts. The seasonal pattern is unchanged, yet the raw output increases mechanically with return scale. A cross-sectional ranking based on unadjusted forecasts may therefore favour more volatile contracts even when they do not contain stronger seasonal information. For this reason, the empirical analysis compares a classical raw-return specification with a volatility-normalised specification.

\paragraph{Volatility normalisation.}
In the volatility-normalised specification, returns are transformed before model estimation. Conditional volatility is estimated separately within each 120-month training window with an exponentially weighted moving average recursion,
\[
\hat \sigma_{i,\tau}^2 = \lambda \hat \sigma_{i,\tau-1}^2 + (1-\lambda) r_{i,\tau-1}^2,
\]
where \(\lambda = 0.97\) for monthly data, following the RiskMetrics recommendation for lower-frequency returns \parencite{JPMorgan1996RiskMetrics}. Standardised returns are then defined as
\[
\tilde r_{i,\tau} = \frac{r_{i,\tau} - \bar r_{i,\tau}}{\hat \sigma_{i,\tau}},
\qquad \tau = 1,\dots,t,
\]
Within a training window, \(\bar r_{i,\tau}\) is the expanding mean from its first observation through \(\tau\). The EWMA variance is warm-started once from the mean squared return of the first 12 observations in that same window; from the second transformed observation onward, it is updated with the lagged-return recursion above. Thus the warm start uses information later than the earliest transformed observations, but neither it nor any subsequent update uses information after the forecast origin \(t\). The resulting score is based entirely on the 120 observations available before the realised trading return at \(t+1\). The transformation changes the interpretation of the signal from predicted raw return to expected return in volatility units, with the aim of improving cross-contract comparability and reducing the mechanical preference for high-volatility contracts. \Textcite{Hassani2020The} motivate the broader point that transformations can affect SSA forecasts in a context-dependent manner; they do not provide the specific EWMA transformation used here.

These procedures yield a one-step-ahead seasonality score for every contract and month under a common rolling framework. The next section translates these scores into trading positions and evaluates their economic usefulness against a common benchmark.

\section{Portfolio Design and Statistical Evaluation}

This section translates the seasonality signals into implementable trading rules and defines the framework used to evaluate their economic and statistical relevance. The guiding principle is comparability. All model classes are estimated on the same rolling windows, transformed into rankings by the same portfolio rule, and judged against the same equal-weight long benchmark under the same cost assumptions. This keeps the comparison focused on a common research question rather than on differences in portfolio engineering.

\subsection{Portfolio formation}

For each commodity contract \(i\) and each evaluation month \(t+1\), the estimation step described above produces a one-step-ahead seasonality score \(s_{i,t+1}\). In the classical specification, this score is the raw fitted seasonal effect or raw forecast produced by DVR, SSA, or RLSSA. In the volatility-normalised specification, the score is obtained from the same model class estimated on volatility-normalised returns. The cross section of contracts is then ranked by these scores each month.

Portfolio construction follows a transparent Top-1 rule. On the long side, the strategy holds the eligible contract with the highest score,
\[
i^{L}_{t+1} = \arg\max_i s_{i,t+1},
\]
and on the short side it holds the eligible contract with the lowest score,
\[
i^{S}_{t+1} = \arg\min_i s_{i,t+1}.
\]

For SSA and RLSSA, every finite forecast is eligible, so these strategies hold one contract every month. These portfolios are relative cross-sectional ranking strategies rather than sign-restricted directional forecast strategies. The long portfolio selects the contract with the highest score even when all available forecasts are negative, while the short portfolio selects the contract with the lowest score even when all available forecasts are positive. The portfolio direction is therefore imposed by the long or short strategy definition, while the forecast determines the relative ranking of contracts. Unlike DVR, SSA and RLSSA have neither an absolute signal-strength threshold nor a sign threshold. The Top-1 rule can consequently force a position when cross-sectional forecast dispersion is economically small or when the selected forecast has a sign opposite to the direction of the portfolio. For DVR, eligibility also requires the sign and Newey--West significance conditions defined above. The portfolio stays in cash without incurring costs when no candidate passes. This deliberately parsimonious design keeps the interpretation of the results straightforward. The Top-1 rule is used as a transparent baseline for economic evaluation, not as a claim that a single-contract portfolio is optimally diversified.

\subsection{Return aggregation}

Monthly returns are implemented on simple returns for compounding. Let \(R_{i,t+1}\) denote the realised simple return of the selected delivery-avoidance contract. Each position is reset to unit notional at the start of the month. The strategy assumes a proportional one-way trading cost \(c\) for each transaction. \Textcite{Marshall2012Commodity} report all-commodity average one-way transaction costs of 6.5 to 8.6 basis points for a \$1{,}000{,}000 trade executed patiently over approximately 60 minutes. The analysis uses 8.6 basis points, the higher of these two aggregate estimates, and sets \(c = 0.00086\). Because a monthly position requires both entry and exit, the one-way cost is applied twice. A short futures position has linear price exposure, so its gross return proxy is \(-R_{i,t+1}\), not the return on an inverse-price asset. Net long and short returns are therefore given by
\[
R^{L,\mathrm{net}}_{t+1} = (1 + R_{i^{L}_{t+1},t+1})(1-c)^2 - 1,
\]
\[
R^{S,\mathrm{net}}_{t+1} = (1 - R_{i^{S}_{t+1},t+1})(1-c)^2 - 1.
\]
These net returns are compounded through time to obtain the long and short portfolio value paths for each model and specification. They are unit-notional futures return proxies: the analysis excludes collateral yield, margin effects, financing, and contract-specific market impact. Months in which DVR remains in cash earn zero by assumption. In particular, the zero-cash convention omits the interest that an uninvested balance could have earned.

\subsection{Benchmark design}

Model performance is evaluated relative to a transparent equal-weight long benchmark. In month \(t+1\), the benchmark holds an equal-weighted long position in all contracts available in the trading universe at that date and applies the same monthly cost convention as the model-based portfolios. If \(N_{t+1}\) denotes the number of eligible contracts in month \(t+1\), its gross return is
\[
R^{EW,L}_{t+1} = \frac{1}{N_{t+1}} \sum_{i=1}^{N_{t+1}} R_{i,t+1},
\]
with the net return obtained after applying the same round-trip cost adjustment.

The equal-weight long portfolio is used for both model directions. It represents the study's low-discretion reference for a conventional long-term commodity strategy and directly answers whether either active long selection or active short selection beats that same baseline. This design is intentional, but its interpretation differs by direction. A long-model comparison primarily combines active contract selection with concentration relative to broad long exposure. A short-model comparison combines the opposite directional exposure with contract selection and therefore must not be interpreted as a pure test of short-side selection skill. A direction-matched equal-weight short portfolio or zero-return cash benchmark would answer a different question and is outside the defined comparison design.

The benchmark is not meant to be an optimal commodity strategy, but a transparent reference portfolio that is easy to interpret and difficult to overfit. Outperformance would indicate improvement relative to this conventional long-only baseline; it would not by itself isolate the separate contributions of direction and contract selection.

\subsection{Bootstrap-based statistical evaluation}

A single historical backtest provides only one realised performance path for each strategy. This is informative, but it does not show how sensitive the results are to alternative return magnitudes around the observed path. The article therefore complements the historical backtest with a path-based Maximum Entropy Bootstrap (MEB) assessment developed by \textcite{VinodLopezdeLacalle2009}. MEB is designed for ordered, potentially nonstationary data and avoids the i.i.d. resampling logic that is not directly valid for dependent time series \parencite{KreissLahiri2012}.

The original MEB first sorts the observed series, generates sorted draws from maximum-entropy intervals formed around its order statistics, and then restores those draws with the original sorting index. Consequently, the temporal ranks of each bootstrap replicate match the ranks of the corresponding observed contract by construction. The procedure perturbs return magnitudes while retaining the rank-based temporal ordering, including the locations of local peaks and troughs and much of the observed periodic structure \parencite{VinodLopezdeLacalle2009}.

This property is central to interpretation in a seasonality study. The MEB paths are alternative magnitude realisations conditional on the observed temporal structure; they are not histories in which the timing of high and low returns is freely resampled. They therefore provide a conditional path-sensitivity assessment for the recursively re-estimated strategies, but not a bootstrap null model that destroys seasonal timing and asks whether seasonality would emerge anew. The phrase ``MEB inference'' below is used in this qualified, conditional sense.

For the results reported here, the procedure uses 500 replications with a fixed random seed of 42, a 10 percent MEB trim, and the core algorithm with boundary reaching enabled. Each contract is bootstrapped separately, and the same simulated 15-contract panel in a given replication is supplied to DVR, SSA, RLSSA, and the equal-weight long benchmark. The design preserves pairing across strategies and the exact within-contract temporal rank order induced by the univariate MEB. It does not explicitly preserve contemporaneous cross-contract dependence, a limitation when interpreting portfolio-level dispersion. All three signal models are fully re-estimated on every MEB path.

The full-period and subperiod tables summarise the distribution of performance over these 500 alternative paths. Cumulative return is the terminal value of compounded net simple returns minus one. The annualised Sharpe ratio uses the arithmetic mean and sample standard deviation of monthly returns, a zero monthly risk-free rate, and \(\sqrt{12}\) annualisation. Maximum drawdown is calculated from a wealth index that includes initial wealth of one. Bootstrap-path means and standard deviations are descriptive sensitivity measures; the 500 paths are not treated as independent observed market histories.

\paragraph{Paired Sharpe assessment.}
The paired assessment is restricted to the Sharpe ratio. For each model and direction, the historical out-of-sample Sharpe difference is evaluated with an approximate two-sided paired, nonstudentized MEB test using the corresponding 500 paired strategy paths. For an observed right-minus-left difference \(d\) and paired bootstrap differences \(d_b^*\), the null distribution is approximated by \(d_b^*-d\). The raw two-sided \(p\)-value is
\[
\widehat p=\frac{1+\sum_{b=1}^{B}\mathbb{I}\!\left(\lvert d_b^*-d\rvert\geq\lvert d\rvert\right)}{B+1},
\qquad B=500,
\]
and the computed diagnostic interval is the corresponding unadjusted 95 percent basic interval. Three comparisons are defined within each direction: volatility-normalised versus classical, classical versus the equal-weight long benchmark, and volatility-normalised versus that benchmark. This produces six families defined by comparison type and direction, with DVR, SSA, and RLSSA tested together within each family. Holm's sequential correction controls the familywise error rate at 5 percent across the three models in each family \parencite{Holm1979}. The add-one correction makes the finite MEB resolution explicit. The pointwise intervals are not used for the Holm decision. A non-rejection is reported as inconclusive and is not interpreted as evidence of equivalence. Because temporal ranks are preserved, even a rejection would be evidence of a conditional performance difference under the observed timing structure, not evidence from a timing-destroying seasonal null. On the short side, any such difference would additionally combine directional exposure and contract selection because the benchmark remains long.

\section{Results}

This section reports the 2016--2024 out-of-sample results. It first presents the realised historical path, followed by full-period descriptive performance across 500 MEB paths and the within-family Holm-adjusted paired Sharpe tests based on the historical effect estimates. It then evaluates subperiod robustness and concludes with an analysis of Top-1 selection stability. All 500 replications are valid in every reported Sharpe comparison. None of the 18 tests rejects after its within-family Holm adjustment; every comparison is therefore reported as inconclusive rather than equivalent.

\subsection{Historical out-of-sample path}

Table~\ref{tab:historical-oos} reports the realised net performance of every strategy on the observed 108-month path from January 2016 through December 2024. All returns use the common transaction-cost convention. CumRet denotes compounded cumulative return and MaxDD denotes maximum drawdown from a wealth index initialised at one. The benchmark is the equal-weight long portfolio used for comparisons with both model directions. The subsequent MEB analysis evaluates how these historical results vary when return magnitudes are perturbed conditional on the observed temporal ranks.

\begin{table}[H]
\centering
\caption{Historical out-of-sample performance, January 2016--December 2024}
\label{tab:historical-oos}
\scriptsize
\renewcommand{\arraystretch}{1.05}

\begin{tabular*}{0.86\textwidth}{@{\extracolsep{\fill}}lllrrr@{}}
\toprule
Model & Direction & Specification & CumRet & Sharpe & MaxDD \\
\midrule
DVR   & Long  & Classical       &  0.2966 &  0.245 & -0.308 \\
DVR   & Long  & Vol.-normalised &  0.2780 &  0.240 & -0.366 \\
DVR   & Short & Classical       & -0.1994 &  0.018 & -0.494 \\
DVR   & Short & Vol.-normalised & -0.0972 &  0.028 & -0.427 \\
SSA   & Long  & Classical       & -0.3241 &  0.014 & -0.605 \\
SSA   & Long  & Vol.-normalised & -0.5776 & -0.112 & -0.812 \\
SSA   & Short & Classical       & -0.3583 &  0.119 & -0.799 \\
SSA   & Short & Vol.-normalised & -0.7548 & -0.205 & -0.867 \\
RLSSA & Long  & Classical       & -0.5764 & -0.088 & -0.721 \\
RLSSA & Long  & Vol.-normalised & -0.7369 & -0.365 & -0.836 \\
RLSSA & Short & Classical       & -0.0054 &  0.184 & -0.825 \\
RLSSA & Short & Vol.-normalised & -0.8125 & -0.288 & -0.917 \\
\addlinespace[0.25em]
Benchmark & Long & Equal-weight & 0.1430 & 0.183 & -0.415 \\
\bottomrule
\end{tabular*}

\end{table}

Historically, both DVR long variants outperform the benchmark in cumulative return and Sharpe ratio, with classical DVR also producing a shallower maximum drawdown. Classical RLSSA short has a Sharpe ratio of 0.184, almost identical to the benchmark's 0.183, but its cumulative return is approximately flat and its maximum drawdown is much deeper at -0.825. The remaining SSA and RLSSA specifications produce negative cumulative returns. These results are descriptive point estimates only. The paired MEB assessment below finds no Holm-adjusted rejection, so the historical differences are not interpreted as robust outperformance.

\subsection{Full-sample performance}

Table~\ref{tab:fullperiod-long} shows that the equal-weight benchmark has the strongest average long-side profile, with a cumulative return of 16.81\%, a Sharpe ratio of 0.191, and a maximum drawdown of -0.414. Among the model portfolios, classical SSA has the highest mean cumulative return at 12.96\% and the highest mean Sharpe ratio at 0.131. However, its median cumulative return is -11.30\%, while its mean maximum drawdown reaches -0.682. This combination indicates that the positive average performance is driven by a subset of favourable MEB paths rather than being representative of the typical path. Classical DVR is approximately flat on average, with a cumulative return of 1.01\%, whereas volatility-normalised DVR loses 5.99\%. Both RLSSA variants are substantially negative. Overall, none of the model specifications matches the benchmark's combination of positive returns, higher risk-adjusted performance, and shallower drawdowns.

\begin{table}[H]
\centering
\caption{Full-period metrics across 500 MEB paths, long side}
\label{tab:fullperiod-long}
\scriptsize
\begin{tabular}{lcccccc}
\toprule
 & \multicolumn{2}{c}{CumRet} & \multicolumn{2}{c}{Sharpe} & \multicolumn{2}{c}{MaxDD} \\
\cmidrule(lr){2-3}\cmidrule(lr){4-5}\cmidrule(lr){6-7}
Strategy & Avg / Median & SD & Avg / Median & SD & Avg / Median & SD \\
\midrule
DVR Classic          & 0.0101 / -0.0202 & 0.2915 & 0.095 / 0.088 & 0.152 & -0.456 / -0.452 & 0.110 \\
DVR Vol.-normalised  & -0.0599 / -0.0779 & 0.2527 & 0.043 / 0.048 & 0.155 & -0.480 / -0.470 & 0.105 \\
SSA Classic          & 0.1296 / -0.1130 & 0.9366 & 0.131 / 0.128 & 0.208 & -0.682 / -0.683 & 0.092 \\
SSA Vol.-normalised  & -0.3183 / -0.4305 & 0.4826 & -0.027 / -0.025 & 0.212 & -0.740 / -0.750 & 0.093 \\
RLSSA Classic        & -0.4900 / -0.6335 & 0.5369 & -0.139 / -0.127 & 0.270 & -0.806 / -0.816 & 0.103 \\
RLSSA Vol.-normalised& -0.6437 / -0.7328 & 0.3069 & -0.287 / -0.267 & 0.274 & -0.820 / -0.844 & 0.106 \\
Equal-weight long benchmark & 0.1681 / 0.1564 & 0.1823 & 0.191 / 0.194 & 0.141 & -0.414 / -0.414 & 0.038 \\
\bottomrule
\end{tabular}
\end{table}

Table~\ref{tab:sig-long} evaluates whether the historical Sharpe differences are supported by the paired MEB inference. Each reported difference is calculated as the first strategy named in the column minus the second, so positive values favour the first strategy. The \(p\)-values are Holm-adjusted across DVR, SSA, and RLSSA within each comparison column.

For DVR, the historical Sharpe ratios are 0.245 for the classical specification and 0.240 for the volatility-normalised specification, compared with 0.183 for the benchmark. These correspond to historical advantages over the benchmark of 0.062 and 0.057. Neither advantage is statistically significant after the paired MEB procedure and Holm adjustment. The historical SSA and RLSSA differences relative to the benchmark are negative, indicating lower historical Sharpe ratios than the benchmark. No long-side comparison rejects the null at the 5\% level. The smallest adjusted \(p\)-value is 0.192 for volatility-normalised RLSSA relative to the benchmark, but the negative Sharpe difference of -0.548 shows that this comparison favours the benchmark rather than RLSSA.

\begin{table}[H]
\centering
\caption{Holm-adjusted paired MEB Sharpe inference, long side}
\label{tab:sig-long}
\scriptsize

\begin{tabular*}{0.88\textwidth}{@{\extracolsep{\fill}}lccc@{}}
\toprule
& \multicolumn{3}{c}{Historical Sharpe difference [Holm-adjusted \(p\)-value]} \\
\cmidrule(lr){2-4}
Model
& \shortstack{Vol.-normalised\\vs.\ Classic}
& \shortstack{Classic\\vs.\ Long BM}
& \shortstack{Vol.-normalised\\vs.\ Long BM} \\
\midrule
DVR
& -0.005 [1.000]
&  0.062 [1.000]
&  0.057 [0.870] \\

SSA
& -0.126 [1.000]
& -0.169 [1.000]
& -0.295 [0.419] \\

RLSSA
& -0.278 [1.000]
& -0.270 [1.000]
& -0.548 [0.192] \\
\bottomrule
\end{tabular*}

\vspace{0.6em}

\begin{minipage}{0.88\textwidth}
\footnotesize
\textit{Notes:} All tests use 500 valid paired MEB paths. No comparison rejects at the 5\% level.
\end{minipage}

\end{table}

On the short side, classical SSA is again the strongest model specification by average Sharpe, but its distribution across MEB paths remains highly asymmetric. Its mean cumulative return is 3.81\%, while its median is -24.75\%. The corresponding mean Sharpe ratio is 0.161 and the mean maximum drawdown is -0.778. Classical RLSSA has a positive mean Sharpe of 0.056, but loses 17.38\% on average and has a mean maximum drawdown of -0.838. Both DVR short portfolios also lose money. Volatility normalisation reduces the average DVR loss from 36.63\% to 19.01\% and improves the mean maximum drawdown from -0.531 to -0.473. In contrast, volatility normalisation materially weakens both SSA-based short portfolios.

The benchmark reported in Table~\ref{tab:fullperiod-short} is intentionally the same equal-weight \emph{long} benchmark used throughout the analysis. The short-side benchmark comparisons should therefore not be interpreted as pure tests of short-contract selection. They also reflect the difference in directional exposure between a short model portfolio and the long-only reference portfolio.

\begin{table}[H]
\centering
\caption{Full-period metrics across 500 MEB paths, short side}
\label{tab:fullperiod-short}
\scriptsize
\begin{tabular}{lcccccc}
\toprule
 & \multicolumn{2}{c}{CumRet} & \multicolumn{2}{c}{Sharpe} & \multicolumn{2}{c}{MaxDD} \\
\cmidrule(lr){2-3}\cmidrule(lr){4-5}\cmidrule(lr){6-7}
Strategy & Avg / Median & SD & Avg / Median & SD & Avg / Median & SD \\
\midrule
DVR Classic          & -0.3663 / -0.3927 & 0.1988 & -0.084 / -0.086 & 0.121 & -0.531 / -0.522 & 0.068 \\
DVR Vol.-normalised  & -0.1901 / -0.2122 & 0.2153 & -0.046 / -0.050 & 0.146 & -0.473 / -0.473 & 0.067 \\
SSA Classic          & 0.0381 / -0.2475 & 0.9221 & 0.161 / 0.150 & 0.192 & -0.778 / -0.799 & 0.113 \\
SSA Vol.-normalised  & -0.6450 / -0.7196 & 0.2972 & -0.191 / -0.195 & 0.185 & -0.846 / -0.863 & 0.083 \\
RLSSA Classic        & -0.1738 / -0.4765 & 1.0262 & 0.056 / 0.047 & 0.237 & -0.838 / -0.873 & 0.115 \\
RLSSA Vol.-normalised& -0.4364 / -0.5764 & 0.5141 & -0.092 / -0.102 & 0.238 & -0.825 / -0.838 & 0.104 \\
Equal-weight long benchmark & 0.1681 / 0.1564 & 0.1823 & 0.191 / 0.194 & 0.141 & -0.414 / -0.414 & 0.038 \\
\bottomrule
\end{tabular}
\end{table}

Table~\ref{tab:sig-short} applies the same paired MEB inference to the short-side Sharpe differences. Classical RLSSA is almost identical to the benchmark in the historical sample, with Sharpe ratios of 0.184 and 0.183 respectively, producing a difference of only 0.002. Classical SSA and DVR have lower historical Sharpe ratios than the benchmark at 0.119 and 0.018. Volatility normalisation also produces negative historical Sharpe differences for SSA and RLSSA relative to their classical specifications.

None of these short-side differences is statistically significant after the Holm adjustment. In particular, the near-zero historical difference between classical RLSSA and the benchmark has an adjusted \(p\)-value of 1.000. The negative differences for volatility-normalised SSA and RLSSA likewise do not provide evidence of statistically robust performance differences after multiplicity adjustment. The short-side results therefore reinforce the descriptive evidence from Table~\ref{tab:fullperiod-short}: individual specifications can appear comparatively stronger on particular metrics, but the paired MEB inference does not support a claim of robust Sharpe outperformance.

\begin{table}[H]
\centering
\caption{Holm-adjusted paired MEB Sharpe inference, short side}
\label{tab:sig-short}
\scriptsize

\begin{tabular*}{0.88\textwidth}{@{\extracolsep{\fill}}lccc@{}}
\toprule
& \multicolumn{3}{c}{Historical Sharpe difference [Holm-adjusted \(p\)-value]} \\
\cmidrule(lr){2-4}
Model
& \shortstack{Vol.-normalised\\vs.\ Classic}
& \shortstack{Classic\\vs.\ Long BM}
& \shortstack{Vol.-normalised\\vs.\ Long BM} \\
\midrule
DVR
&  0.010 [0.960]
& -0.164 [1.000]
& -0.154 [0.533] \\

SSA
& -0.324 [0.437]
& -0.064 [1.000]
& -0.388 [0.395] \\

RLSSA
& -0.472 [0.631]
&  0.002 [1.000]
& -0.470 [0.407] \\
\bottomrule
\end{tabular*}

\vspace{0.6em}

\begin{minipage}{0.88\textwidth}
\footnotesize
\textit{Notes:} All tests use 500 valid paired MEB paths. No comparison rejects at the 5\% level.
\end{minipage}

\end{table}

Full-period results provide the broad ranking of the strategies, but they do not establish whether these relative outcomes persist across different market environments. The analysis therefore next divides the out-of-sample period into three non-overlapping subperiods to examine the stability of the observed performance differences.

\subsection{Subperiod robustness}

To examine whether the full-sample results are stable across different market environments, the 2016--2024 out-of-sample period is split into three non-overlapping subperiods: 2016 to 2018, 2019 to 2021, and 2022 to 2024. Tables~\ref{tab:sub_long_appendix} and~\ref{tab:sub_short_appendix} report the corresponding subperiod means and show that the model rankings are not stable across market regimes.

On the long side, no model strategy is positive in all three subperiods. DVR is weak in the 2019--2021 window, while SSA is sharply negative in 2016--2018 but strong in 2019--2021 and remains positive in 2022--2024. Classical RLSSA is negative in two of the three windows, and volatility-normalised RLSSA is negative in all three. The benchmark is also negative in the first subperiod, but it is positive in the two later windows and has a more stable full-period profile than the model portfolios.

On the short side, the strongest model episodes are concentrated in particular windows rather than spread evenly across the sample. DVR short performs well in 2019--2021 but loses money in the first and final subperiods. Classical SSA and RLSSA are strongest in 2022--2024, while their earlier-window performance is weak or negative. The equal-weight long benchmark is negative in 2016--2018 and positive in both later windows. The later sample includes COVID-19 and the Russia--Ukraine war, two episodes that are associated in the literature with unusual oil-market dynamics and broad commodity supply disruptions \parencite{Le2021The, WorldBank2022CommodityOutlook}. These references do not prove that the strategies profit from crisis timing. Rather, they support the narrower interpretation that the later subperiods occurred in a less stable commodity-market environment. The results suggest that the seasonal signals are regime-sensitive and that the full-period findings are not driven by a stable positive premium that persists through all market environments.

The subperiod tables reinforce the same central conclusion as the full-sample results: positive performance is localised in particular combinations of models and periods and is too uneven to support a broad claim of robust out-of-sample outperformance.

Temporal robustness is only one dimension of stability. A second question is whether the classical and volatility-normalised specifications actually produce similar trades, or whether they arrive at their results through materially different contract selections.

\subsection{Ticker robustness and path differentiation}

Table~\ref{tab:ticker-robustness} compares the leading Top-1 selections of the classical and volatility-normalised specifications across the 500 MEB paths. Avg.\ C1 and Avg.\ R1 report the average path share of the most frequently selected traded contract for the classical and volatility-normalised variants. For DVR, C cash and V cash report the average share of paths without a trade. Match is the percentage of months in which both variants select the same leading traded ticker, conditional on both having a traded Top-1 entry, while \(n\) gives the number of comparable months. SSA and RLSSA always trade, so all 108 months are comparable for these models.

\begin{table}[H]
\centering
\caption{Top-1 selection stability across 500 MEB paths}
\label{tab:ticker-robustness}
\scriptsize
\setlength{\tabcolsep}{6pt}
\renewcommand{\arraystretch}{1.05}

\begin{tabular}{llrrrrrr}
\toprule
Model & Side
& Avg.\ C1 (\%)
& Avg.\ R1 (\%)
& C cash (\%)
& V cash (\%)
& Match (\%)
& \(n\) \\
\midrule
DVR   & Long  & 64.55 & 65.17 & 49.54 & 52.09 & 61.54 & 65  \\
DVR   & Short & 77.49 & 72.78 & 47.56 & 48.20 & 73.02 & 63  \\
SSA   & Long  & 66.82 & 70.66 & --    & --    & 50.00 & 108 \\
SSA   & Short & 66.60 & 69.90 & --    & --    & 52.78 & 108 \\
RLSSA & Long  & 52.86 & 53.29 & --    & --    & 59.26 & 108 \\
RLSSA & Short & 54.90 & 54.48 & --    & --    & 52.78 & 108 \\
\bottomrule
\end{tabular}

\end{table}

The SSA-based Top-1 choices are only moderately concentrated, with leading-contract shares ranging from approximately 53\% to 71\% and classical versus volatility-normalised match rates between 50\% and 59\%. DVR selections are more concentrated, particularly on the short side, but roughly half of the paths remain in cash on average. On the long side, 26 months are flat in all 500 paths under both DVR variants, with seven additional months lacking a classical traded leader and ten lacking a volatility-normalised traded leader, leaving 65 comparable months. The corresponding short-side counts are 30, nine, and six, leaving 63 comparable months. The DVR filter therefore changes both market exposure and contract selection.

\begin{figure}[H]
\centering
\includegraphics[width=0.32\linewidth]{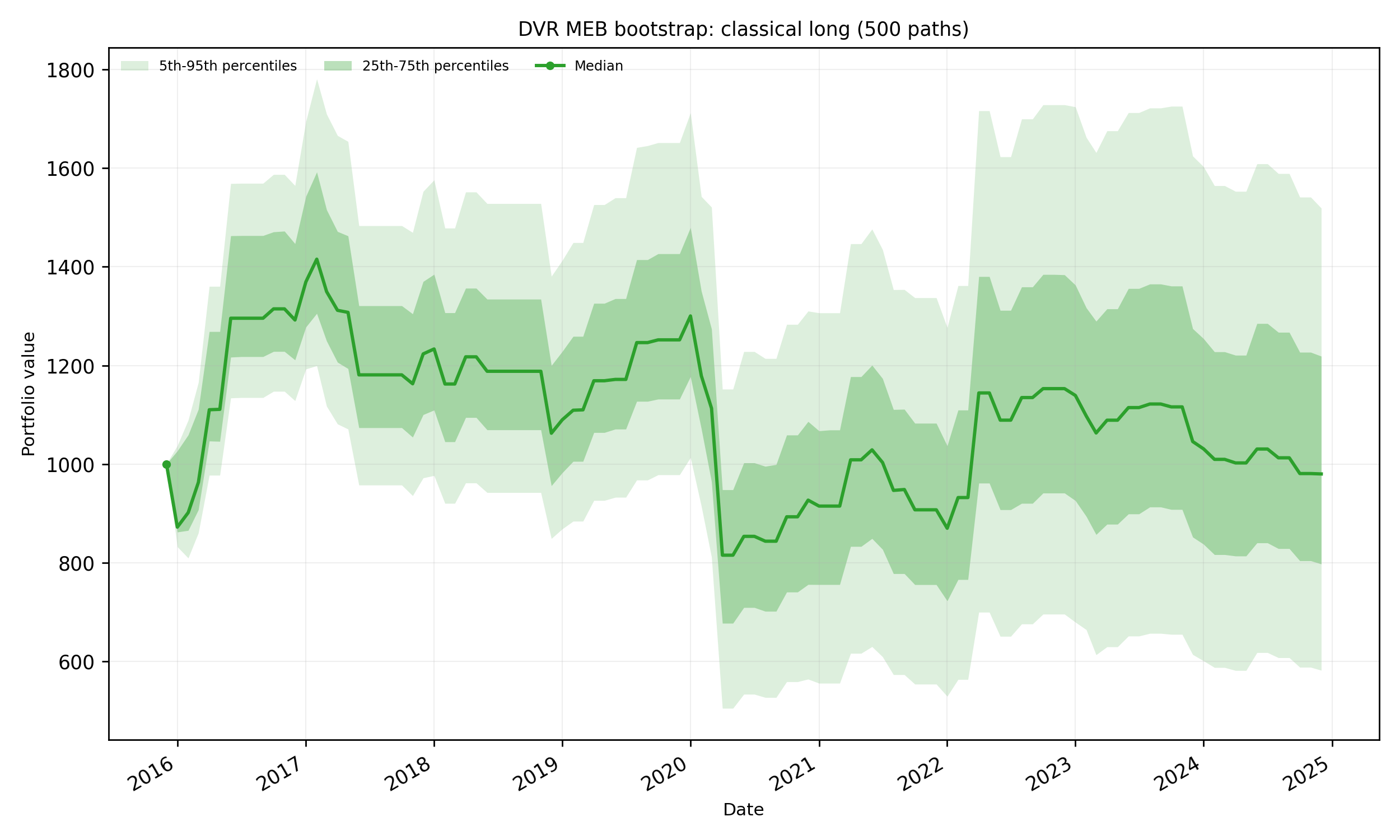}\hfill
\includegraphics[width=0.32\linewidth]{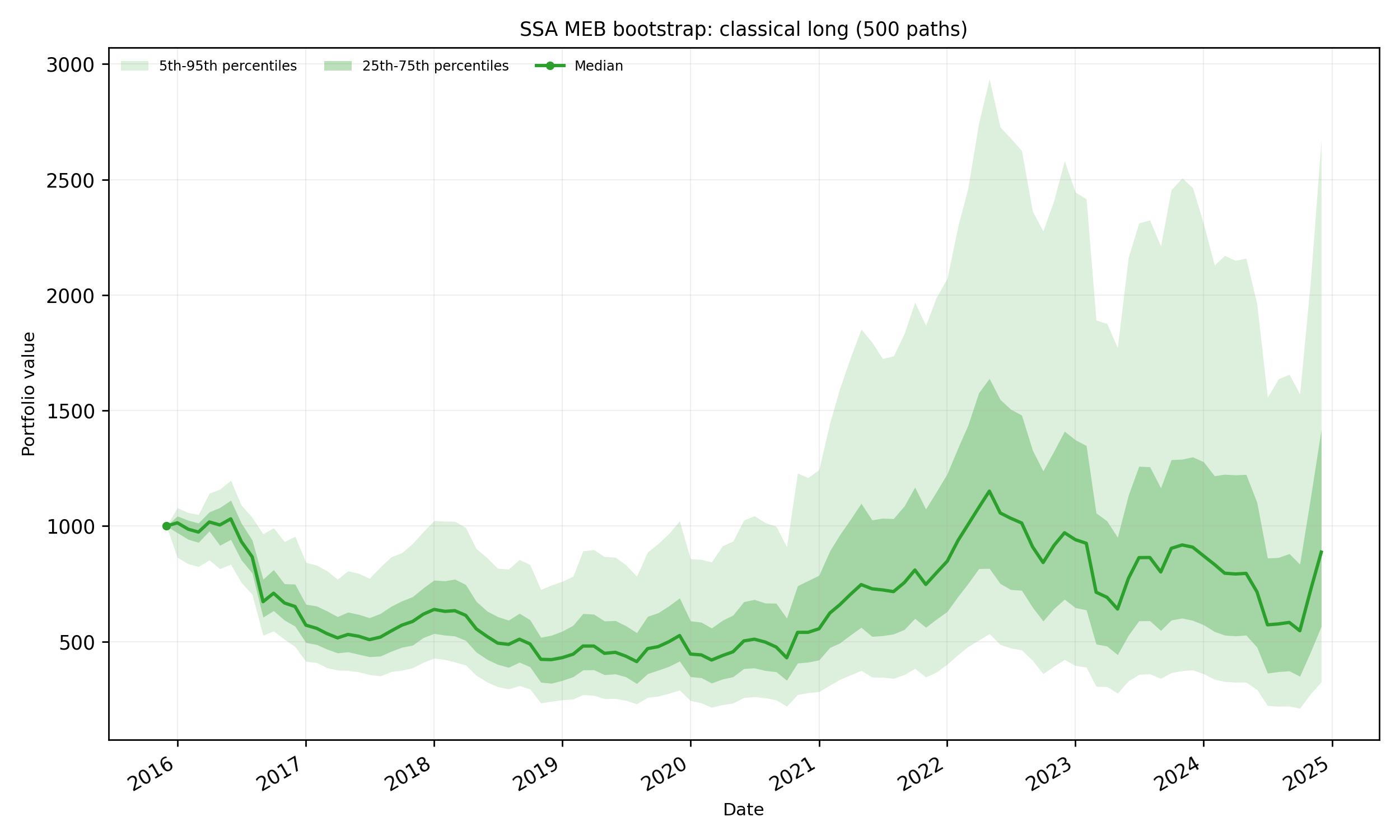}\hfill
\includegraphics[width=0.32\linewidth]{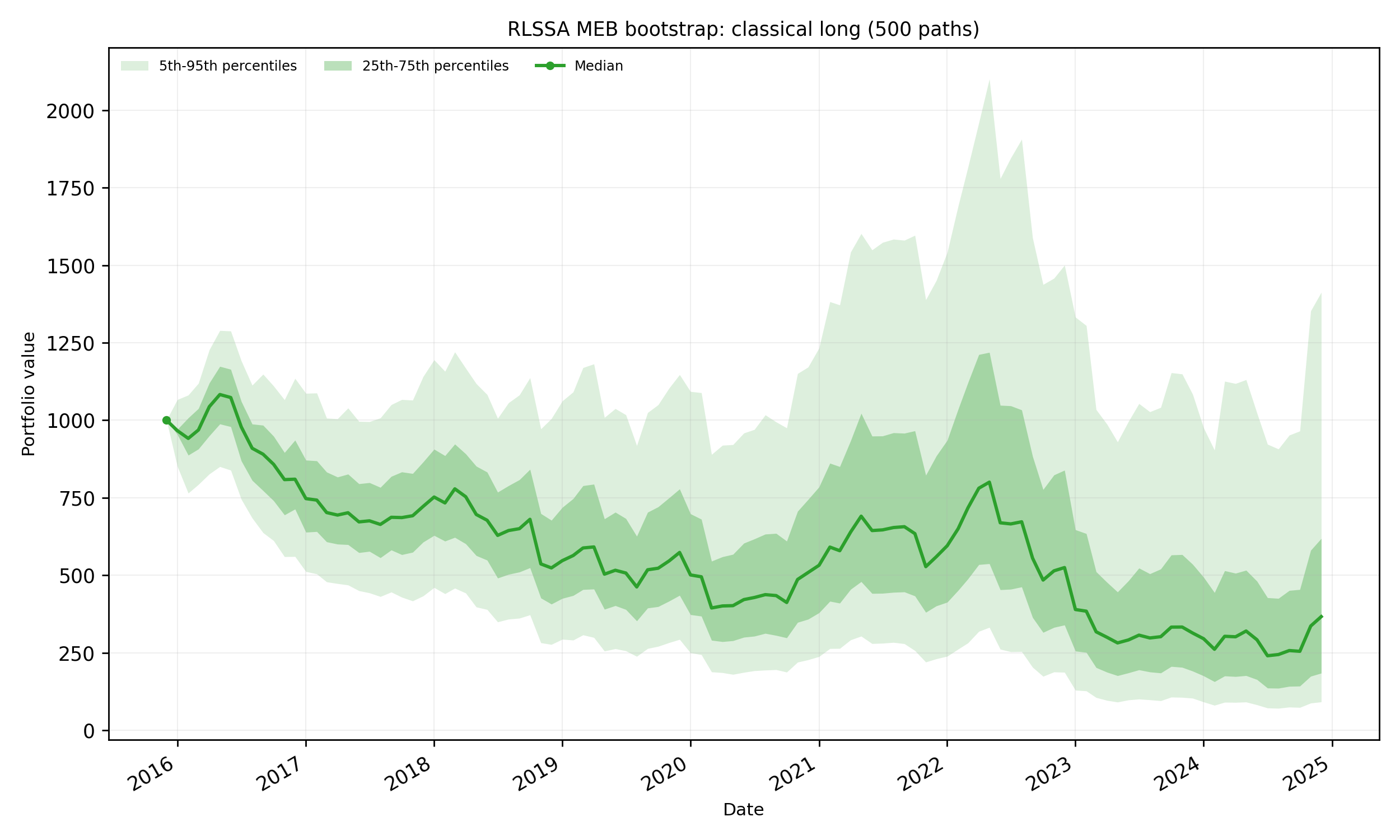}\par\vspace{0.5em}
\includegraphics[width=0.32\linewidth]{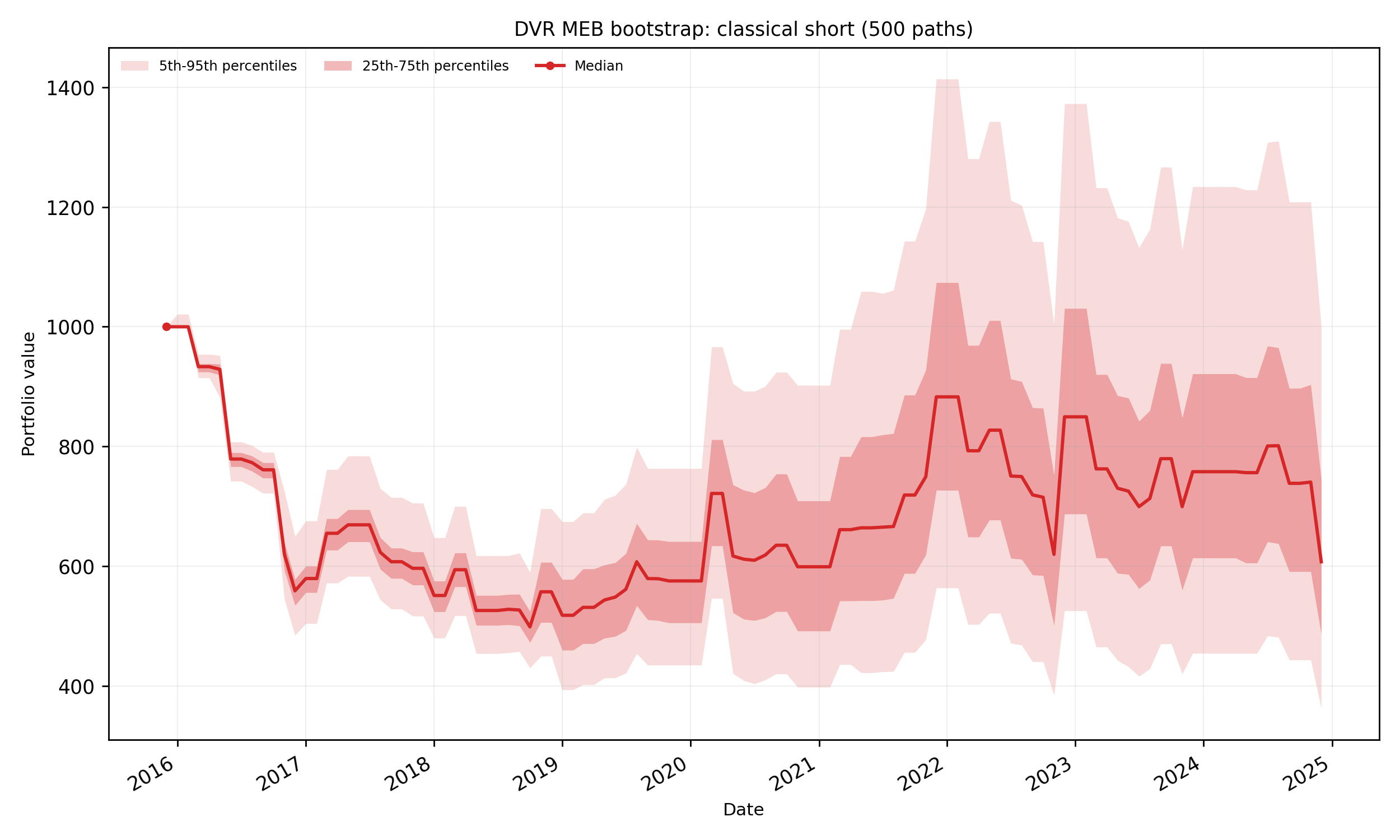}\hfill
\includegraphics[width=0.32\linewidth]{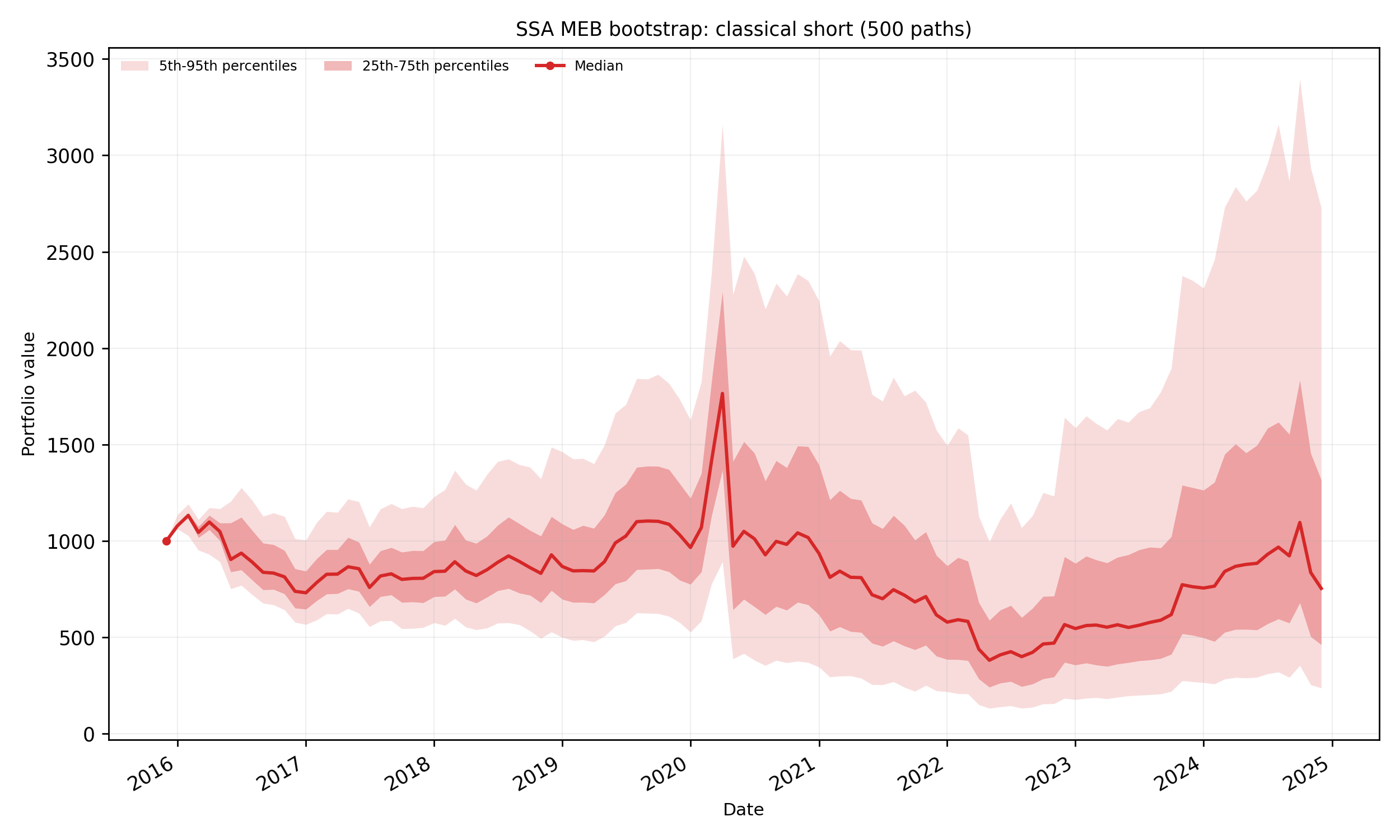}\hfill
\includegraphics[width=0.32\linewidth]{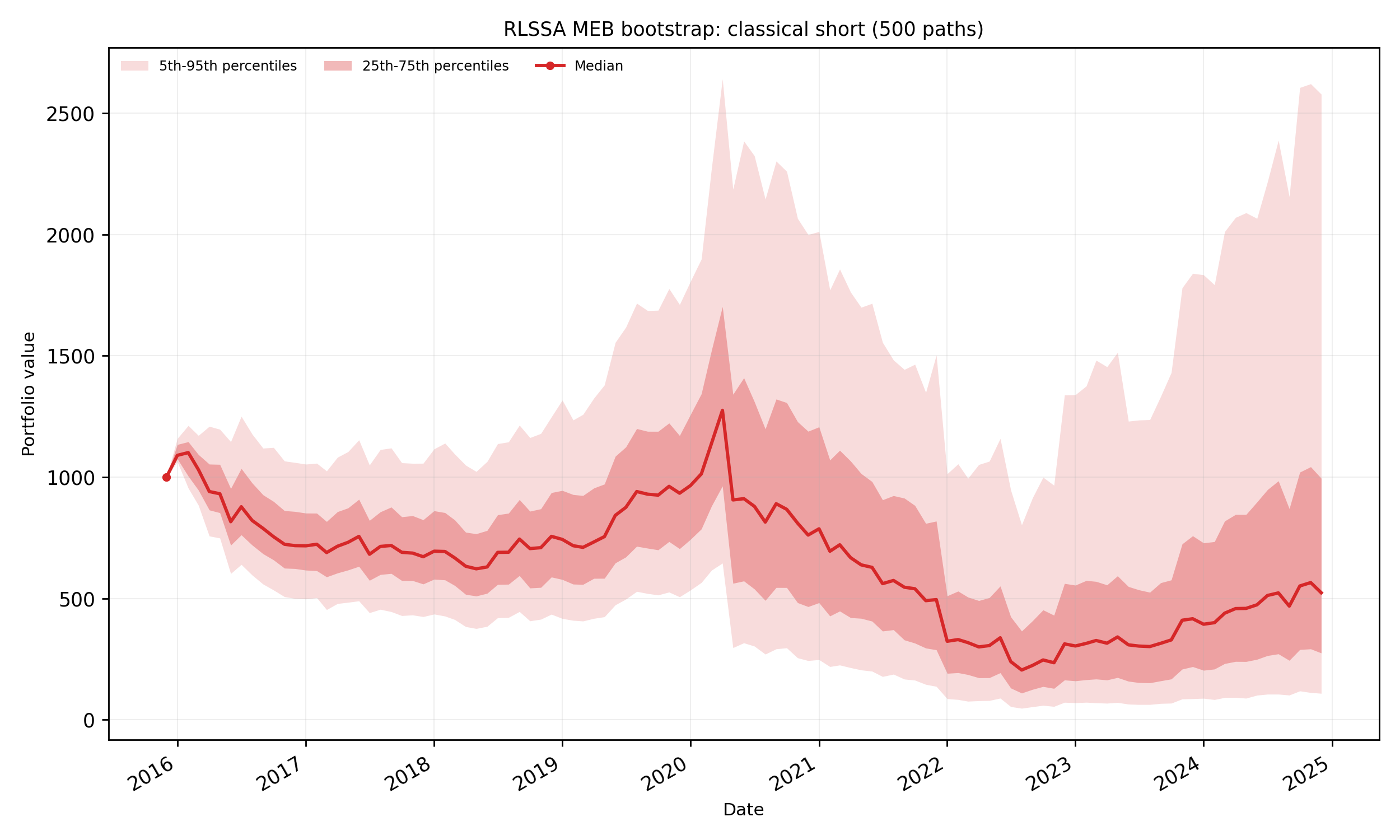}
\caption{Pointwise median cumulative portfolio values with central 50\% and 90\% bands across 500 MEB replications for the classical DVR, SSA, and RLSSA strategies. The top row shows long portfolios and the bottom row shows short portfolios.}
\label{fig:bootstrap_paths}
\end{figure}

Volatility normalisation also frequently changes the selected contract within the same model class. Appendix Figure~\ref{fig:bootstrap_paths} complements this evidence with pointwise medians and central bands across the 500 classical long and short replications. The SSA-based strategies exhibit deep drawdowns, while RLSSA short shows the greatest terminal-return dispersion among the reported strategies. Together, the selection and path evidence indicate that both normalisation and trade direction materially affect implementation risk in a model-specific way.

The results do not identify a simple empirical winner and instead motivate the subsequent discussion of why seasonal signals appear in particular combinations of models and periods without producing stable full-period trading premia.

\section{Discussion}

The results contribute to the commodity-seasonality literature by connecting existing evidence on return seasonality with an implementation-aware comparison of alternative signal models. \Textcite{KeloharjuLinnainmaaNyberg2016} document recurring same-calendar-month return seasonalities in commodities, while \textcite{LiLiuMiaoTse2024} show that commodity-futures seasonal effects have weakened in more recent periods. Against this background, the present finding that none of the model portfolios establishes robust benchmark outperformance from 2016 to 2024 is broadly consistent with the view that recent seasonal effects are economically fragile, while extending the literature through a direct comparison of DVR, SSA, and RLSSA under common rolling estimation, transaction-cost, and benchmark rules. Relative to \textcite{DEGENHARDT2018169}, the analysis considers a broader contract-level and month-by-month setting, while extending the SSA and RLSSA literature from signal extraction and forecasting toward common portfolio rules, transaction costs, and benchmark evaluation \parencite{Rodrigues2018RobustSSA, KazemiRodrigues2025}. The most informative findings concern changes in model performance across market environments and the selective effect of volatility normalisation.

\subsection{Model performance across market environments and the role of uncertainty}

The most interesting subperiod result is not that one seasonality model dominates throughout the sample, but that the relative ranking of the models changes with the market environment. On the long side, DVR performs comparatively well in 2016--2018, while both SSA-based approaches are weak. This pattern reverses in 2019--2021, when classical SSA becomes the strongest model and classical RLSSA also moves into positive territory. The evidence points toward model performance that varies materially across the observed market subperiods rather than a universally superior representation of seasonality.

One possible explanation lies in how the models respond to changes in seasonal structure. \Textcite{TangEtAl2024} show that both the amplitude and periodic structure of commodity seasonality can vary during periods of elevated market uncertainty. The later subperiods in the present study also contain economically exceptional episodes. COVID-19 generated historically unusual oil-market dynamics \parencite{Le2021The}, while the Russia--Ukraine war produced major disruptions across global commodity markets \parencite{WorldBank2022CommodityOutlook}. These events do not establish why any particular strategy performed well, but they confirm that the later windows represent substantially different market environments from 2016--2018. Uncertainty may therefore alter the seasonal structure that the models attempt to recover rather than merely adding noise to an otherwise stable process.

One possible explanation for the shift from DVR toward SSA in 2019--2021 is the greater flexibility of SSA. DVR searches for comparatively stable calendar-month effects and trades only when its eligibility conditions are satisfied. Such a restrictive specification may be advantageous when recurring monthly relationships remain relatively stable. SSA instead allows several oscillatory components to contribute to the signal without imposing a single fixed seasonal frequency. If uncertainty changes the amplitude or periodicity of recurring commodity patterns, the additional flexibility of SSA may become more valuable. A natural extension would be to condition model choice on the market state. \Textcite{BahloulEtAl2018} show that measures including the VIX and economic uncertainty contain nonlinear predictive information for commodity futures returns and volatility, while \textcite{GuidolinPedio2021} find that the usefulness of commodity predictors can differ across volatility states. A future strategy could therefore estimate an uncertainty state using only information available at the forecast origin and switch between DVR and SSA-based signals. The VIX is one possible indicator, although commodity-specific volatility or broader uncertainty measures may provide a more appropriate state variable.

A second finding requiring explanation is that classical SSA performs better than RLSSA during the strongest SSA subperiod. Robust SSA is designed to reduce the influence of observations treated as contamination \parencite{Rodrigues2018RobustSSA, KazemiRodrigues2025}. In commodity markets, however, an extreme observation during a disruption is not necessarily irrelevant contamination. Large movements can reflect genuine changes in supply, inventories, demand, or geopolitical conditions. Reducing their influence may therefore remove economically meaningful information together with noise. The robust-SSA literature also does not imply that an \(\ell_1\)-based decomposition should universally dominate classical SSA in forecasting. Robust signal extraction and superior trading performance remain distinct objectives.

At the same time, the exceptionally strong SSA performance in 2019--2021 should not automatically be interpreted as evidence that flexible seasonal models benefit from uncertainty. Commodity futures are characterised by infrequent but very large price movements \parencite{NguyenProkopczuk2019}. Under the Top-1 design used here, favourable exposure to only a few extreme realised returns can have a disproportionate effect on compounded subperiod performance. The strong SSA result could therefore reflect genuine adaptation to changing seasonal structure, favourable exposure to unusually large price movements, or a combination of both. Distinguishing these mechanisms is important before concluding that uncertainty systematically favours one model class.

\subsection{Why volatility normalisation helps only selectively}

A second dimension of model dependence concerns volatility normalisation. Ex ante, the transformation is well motivated because raw seasonality scores inherit the scale of the underlying return series. Ranking unadjusted signals can therefore mechanically favour high-volatility contracts even when their underlying seasonal structure is not stronger. The scaling experiment illustrates this problem directly. Yet the portfolio results show that improved comparability of signals does not translate into uniformly better trading outcomes.

This pattern is only partially comparable to \textcite{Hassani2020The}. For SSA applied to monthly data, they show that standardisation can improve forecast accuracy at horizons that include (h=1), but they do not find a general improvement from transformation across all settings. Their analysis concerns forecast accuracy for individual time series, whereas the present study uses volatility-normalised forecasts to rank contracts cross-sectionally and form trading portfolios. The distinction matters because a transformation that improves statistical comparability need not improve the economic ranking of assets.

The empirical results illustrate this trade-off. Volatility normalisation generally weakens the SSA and RLSSA portfolios in the full-period analysis, while for DVR short it reduces the average loss without producing statistically conclusive benchmark outperformance. Once the problem shifts from univariate forecasting to portfolio selection, return amplitude may contain economic information rather than representing only an undesirable scaling effect. Commodity futures prices are connected to storage conditions and convenience yields, while futures premia are also related to producers' hedging demand and the risk-bearing capacity of speculators \parencite{LIU20101675, ACHARYA2013441}. Large raw signals may therefore partly reflect economically meaningful differences across commodity markets.

The mixed results can consequently be interpreted as a trade-off. Volatility normalisation is useful when the gain in cross-contract comparability outweighs the information lost by suppressing differences in return amplitude. When amplitude itself contains information relevant for ranking future opportunities, the transformation may weaken rather than improve the portfolio signal. This helps explain why volatility normalisation has no uniform effect across DVR, SSA, and RLSSA and cautions against treating normalisation as an automatically superior preprocessing step.

\subsection{Future research}

The regime-dependent results suggest a first and particularly important extension. Future work could construct an ex ante state-dependent strategy in which observable uncertainty determines which seasonality model is applied. The switching rule would need to rely exclusively on information available before the trading month and should be specified before evaluating subsequent returns. Its performance could then be compared with the fixed-model strategies under the same transaction-cost and benchmark framework. To determine whether any apparent advantage is genuinely related to changing seasonal structure, the analysis should also decompose returns by month and contract and examine whether the result survives the removal or truncation of the largest positive and negative observations. This would distinguish persistent state dependence from performance dominated by a small number of extreme commodity moves.

A second extension concerns the treatment of weak signals and model specification. SSA and RLSSA currently enter a position whenever a finite forecast is available, whereas DVR can remain in cash when its eligibility conditions are not satisfied. Introducing an explicit signal-strength threshold for the SSA-based strategies would test whether some of their weak full-period performance results from forcing positions when seasonal information is economically small. A separately preregistered extension could also examine contract-specific rank selection rather than the common (r=12) specification used here \parencite{Rodrigues2018RobustSSA, KazemiRodrigues2025}. Both changes should be treated as new specifications rather than retrospective tuning of the current results.

A third extension concerns the statistical evaluation. The present MEB is applied separately to each contract and therefore does not explicitly preserve contemporaneous cross-contract dependence. A multivariate resampling procedure could provide a stronger assessment of portfolio-level dispersion. A timing-randomising null would answer a complementary question by testing the strategies after deliberately removing seasonal timing. These extensions would help distinguish sensitivity to return magnitudes, dependence across commodity markets, and the existence of seasonal timing itself.

\section{Conclusion}

This article examines whether seasonality in commodity futures survives an out-of-sample evaluation that incorporates liquidity constraints, transaction costs, and path-based robustness assessment. Across 500 MEB paths, the equal-weight long benchmark has the strongest average full-period return and Sharpe ratio and the least severe average maximum drawdown. None of the 18 paired MEB Sharpe tests rejects after the within-family Holm adjustment. These non-rejections are treated as inconclusive rather than as evidence of equivalence and occur alongside negative model medians, deep drawdowns, and substantial variation across subperiods.

The results support three main conclusions. First, seasonal forecasts that remain visible under rolling estimation do not translate into statistically robust outperformance of the equal-weight long benchmark under the present trading design. Second, the usefulness of the signals is strongly model-dependent and varies materially across market subperiods. DVR performs comparatively well in the earlier long-side subperiod, whereas the strongest SSA-based results emerge in the later windows that contain major commodity-market disruptions. This suggests that the relevant representation of seasonality may change with the market environment rather than remain constant through time. Third, neither methodological refinement produces a uniform improvement. RLSSA does not consistently outperform classical SSA, and volatility normalisation improves some outcomes while weakening others.

The main contribution is not the identification of a single exploitable seasonal premium, but evidence that the economic value of commodity seasonality depends on how the signal is represented and on the environment in which it is traded. The subperiod results point toward a natural next step in which observable uncertainty measures are used ex ante to determine which seasonal model is applied, combined with explicit tests of whether apparent regime advantages survive the removal of extreme commodity price movements. Such an approach would move beyond asking whether seasonality works on average and instead test when particular representations of seasonality are economically useful. The public repository provides code and derived monthly panels, while the complete 500-run artifacts are available from the authors subject to the original data provider's access conditions.

\newpage

\phantomsection
\section*{Appendices}
\markboth{APPENDICES}{APPENDICES}

\begin{table}[H]
\centering
\caption{Sample volume by futures contract, January 2001--December 2024}
\label{tab:sample-volume}
\scriptsize
\setlength{\tabcolsep}{4pt}
\begin{tabular}{llrr}
\toprule
Ticker & Contract & Mean Avg. Daily Vol. & Min. Avg. Daily Vol. \\
\midrule
CC & Cocoa                & 13{,}633  & 2{,}857  \\
CF & Coffee               & 15{,}205  & 4{,}063  \\
CO & WTI Crude Oil        & 224{,}915 & 48{,}893 \\
CP & Copper (High Grade)  & 33{,}015  & 1{,}007  \\
CT & Cotton \#2           & 13{,}669  & 1{,}061  \\
GD & Gold                 & 86{,}320  & 1{,}084  \\
HE & Lean Hogs (Globex)   & 14{,}121  & 2{,}182  \\
HO & Heating Oil          & 35{,}576  & 9{,}283  \\
LE & Live Cattle (Globex) & 18{,}534  & 3{,}970  \\
NG & Natural Gas          & 70{,}539  & 9{,}241  \\
SU & Sugar \#11           & 47{,}906  & 8{,}101  \\
SV & Silver 5000 oz       & 32{,}004  & 1{,}531  \\
ZC & Corn (Globex)        & 115{,}990 & 11{,}093 \\
ZS & Soybeans (Globex)    & 70{,}199  & 4{,}537  \\
ZW & Wheat (Globex)       & 46{,}260  & 6{,}957  \\
\bottomrule
\end{tabular}
\end{table}

\begin{table}[H]
\centering
\caption{Subperiod means across 500 MEB paths, long side}
\label{tab:sub_long_appendix}
\scriptsize
\setlength{\tabcolsep}{4pt}
\begin{tabular}{llrrrrrrrr}
\toprule
 & & \multicolumn{3}{c}{Classic} & \multicolumn{3}{c}{Vol.-normalised} & \multicolumn{2}{c}{Long benchmark} \\
\cmidrule(lr){3-5}\cmidrule(lr){6-8}\cmidrule(lr){9-10}
Model & Window & CumRet & Sharpe & MaxDD & CumRet & Sharpe & MaxDD & CumRet & Sharpe \\
\midrule
DVR & 2016 to 2018 & 0.0816 & 0.217 & -0.258 & -0.0629 & -0.066 & -0.244 & -0.1355 & -0.534 \\
DVR & 2019 to 2021 & -0.1489 & -0.110 & -0.392 & -0.1542 & -0.146 & -0.386 & 0.1922 & 0.463 \\
DVR & 2022 to 2024 & 0.0966 & 0.230 & -0.190 & 0.1827 & 0.339 & -0.225 & 0.1251 & 0.370 \\
SSA & 2016 to 2018 & -0.5548 & -0.937 & -0.621 & -0.6230 & -1.078 & -0.678 & -0.1355 & -0.534 \\
SSA & 2019 to 2021 & 1.0959 & 0.861 & -0.273 & 0.5614 & 0.610 & -0.285 & 0.1922 & 0.463 \\
SSA & 2022 to 2024 & 0.1906 & 0.240 & -0.562 & 0.1347 & 0.223 & -0.570 & 0.1251 & 0.370 \\
RLSSA & 2016 to 2018 & -0.4348 & -0.660 & -0.549 & -0.3812 & -0.585 & -0.517 & -0.1355 & -0.534 \\
RLSSA & 2019 to 2021 & 0.1938 & 0.281 & -0.407 & -0.0895 & -0.051 & -0.410 & 0.1922 & 0.463 \\
RLSSA & 2022 to 2024 & -0.2864 & -0.166 & -0.727 & -0.3652 & -0.309 & -0.694 & 0.1251 & 0.370 \\
\bottomrule
\end{tabular}
\end{table}

\begin{table}[H]
\centering
\caption{Subperiod means across 500 MEB paths, short side}
\label{tab:sub_short_appendix}
\scriptsize
\setlength{\tabcolsep}{4pt}
\begin{tabular}{llrrrrrrrr}
\toprule
 & & \multicolumn{3}{c}{Classic} & \multicolumn{3}{c}{Vol.-normalised} & \multicolumn{2}{c}{Long benchmark} \\
\cmidrule(lr){3-5}\cmidrule(lr){6-8}\cmidrule(lr){9-10}
Model & Window & CumRet & Sharpe & MaxDD & CumRet & Sharpe & MaxDD & CumRet & Sharpe \\
\midrule
DVR & 2016 to 2018 & -0.4408 & -0.771 & -0.499 & -0.1928 & -0.324 & -0.262 & -0.1355 & -0.534 \\
DVR & 2019 to 2021 & 0.6356 & 0.797 & -0.212 & 0.6848 & 0.956 & -0.135 & 0.1922 & 0.463 \\
DVR & 2022 to 2024 & -0.3117 & -0.280 & -0.334 & -0.4066 & -0.876 & -0.408 & 0.1251 & 0.370 \\
SSA & 2016 to 2018 & -0.0481 & 0.008 & -0.407 & -0.2081 & -0.271 & -0.427 & -0.1355 & -0.534 \\
SSA & 2019 to 2021 & -0.2385 & 0.028 & -0.640 & -0.6116 & -0.538 & -0.748 & 0.1922 & 0.463 \\
SSA & 2022 to 2024 & 0.4298 & 0.435 & -0.452 & 0.1234 & 0.259 & -0.415 & 0.1251 & 0.370 \\
RLSSA & 2016 to 2018 & -0.2140 & -0.259 & -0.474 & -0.1537 & -0.169 & -0.432 & -0.1355 & -0.534 \\
RLSSA & 2019 to 2021 & -0.2134 & -0.087 & -0.635 & -0.4737 & -0.527 & -0.670 & 0.1922 & 0.463 \\
RLSSA & 2022 to 2024 & 0.2832 & 0.367 & -0.571 & 0.2624 & 0.337 & -0.473 & 0.1251 & 0.370 \\
\bottomrule
\end{tabular}
\end{table}

\clearpage
\printbibliography
\end{document}